\documentclass[11pt]{article}
\usepackage{epsfig}
\usepackage{amsmath, amssymb}
\begin{document}
\newcommand{\Eins}{{\mathbf 1}} 
\newcommand{\RR}{{\mathbb R}}
\newcommand{\NN}{{\mathbb N}}
\newcommand{\ZZ}{{\mathbb Z}}
\newcommand{\CC}{{\mathbb C}} 
\newcommand{\HH}{{\cal H}} 
\newcommand{\UU}{{\cal U}}
\renewcommand{\SS}{{\cal S}}
\newcommand{\XX}{{\cal X}} 
\newcommand{\BB}{{\cal B}} 
\newcommand{\be}{\begin{equation}} 
\newcommand{\ee}{\end{equation}}
\newcommand{\bea}{\begin{eqnarray}} 
\newcommand{\ea}{\end{eqnarray}}
\renewcommand{\theequation}{\thesection.\arabic{equation}}
\newcommand{\QED}{\hspace*{\fill}Q.E.D.\vskip2mm}
\newcommand{\End}{\hbox{End}} 
\newcommand{\Hom}{\hbox{Hom}} 
\newcommand{\id}{\hbox{id}} 
\newcommand{\Tr}{\hbox{Tr\,}} 
\newcommand{\Ad}{\hbox{Ad}} 
\newcommand{\eps}{\varepsilon} 
\newcommand{\ind}{^{\rm ind}} 
\newcommand{\dual}{^{\rm dual}} 
\newcommand{\inv}{^{-1}} 
\newcommand{\opp}{^{\rm opp}} 
\newcommand{\gen}{^{\rm gen}}  
\renewcommand{\max}{^{\rm max}} 
\newcommand{\rest}{\!\restriction}
\title{\bf Local Fields in Boundary Conformal QFT}  
\author{{\sc Roberto
Longo} \\ Dipartimento di Matematica, Universit\`a di Roma ``Tor
Vergata'', \\ 00133 Roma, Italy\thanks{supported in part by GNAMPA-INDAM, MIUR
and EU-HPP; electronic address: 
{\tt longo@mat.uniroma2.it}} \\ and \\[4mm] {\sc Karl-Henning Rehren} \\
Institut f\"ur Theoretische Physik, Universit\"at G\"ottingen,
\\ 37077 G\"ottingen, Germany\thanks{Electronic address:
{\tt rehren@theorie.physik.uni-goe.de}}\\[5mm]{\sl Dedicated to
Detlev Buchholz on the occasion of his 60th birthday}  } 

\maketitle

\begin{abstract} Conformal field theory on the half-space $x>0$ of
  Minkowski space-time (``boundary CFT'') is analyzed from an algebraic 
  point of view, clarifying in particular the algebraic structure of 
  local algebras and the bi-localized charge structure of local
  fields. The field content and the admissible boundary conditions are 
  characterized in terms of a non-local chiral field algebra.
\end{abstract}

\vskip15mm 
PACS 2003: 03.70.+k

MSC 2000: 81R15, 81T05, 81T40

\newpage

\section{Introduction}
\setcounter{equation}{0} 

We study relativistic boundary CFT on the half-plane $M_+=\{(t,x): x>0\}$. 
This is a local QFT with a conserved and traceless stress-energy tensor, 
subject to a boundary condition at the boundary $x=0$. As is well known, 
conservation and vanishing of the trace imply that the components 
$T_L=\frac12(T_{00}+ T_{01})$ and $T_R=\frac12(T_{00}- T_{01})$ are
chiral fields, $T_L=T_L(t+x)$, $T_R=T_R(t-x)$.
The boundary condition is the absence of energy flow across the boundary, 
\begin{equation} 
T_{01}(t,x=0) = 0 \qquad\Leftrightarrow \qquad T_L=T_R\equiv T. 
\end{equation}
It follows that the components $T_{10}=T_{01}$, $T_{11}=T_{00}$ of the
stress-energy tensor are of the form 
\begin{equation} 
T_{00}(t,x) = T(t+x) + T(t-x), \qquad T_{01}(t,x) = T(t+x) - T(t-x), 
\end{equation}
i.e., {\em bi-local} expressions in terms of the chiral field $T$
(cf.\ Fig.\ 1).

\begin{minipage}{115mm}\hskip40mm \epsfig{file=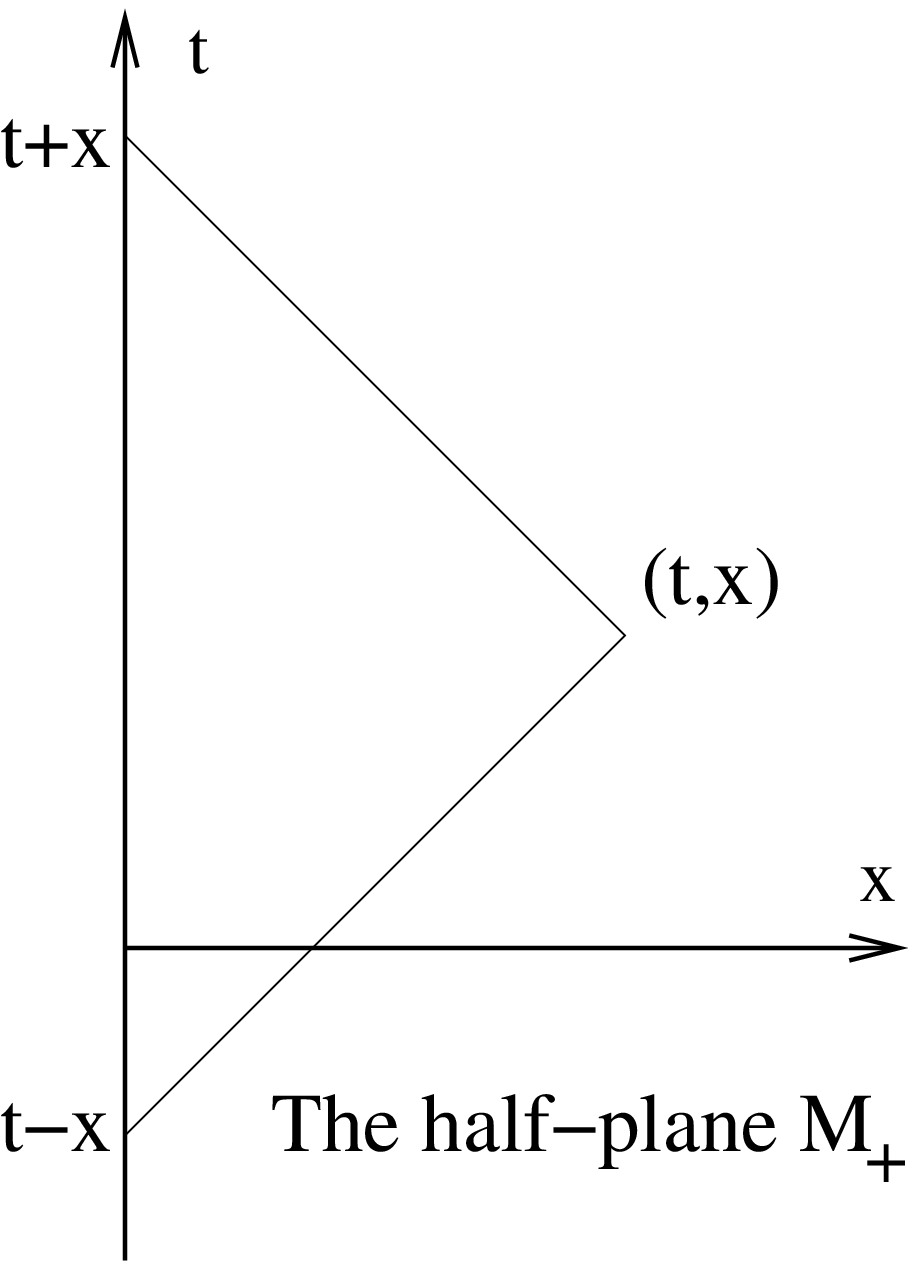,width=3.5cm} 

\small {\bf Fig.\ 1:} A point in the half space $M_+$. A canonical field
localized at $(t,x)$ is a bi-local linear combination of chiral field
localized at $t+x$ and $t-x$.
\end{minipage}
\vskip3mm
Apart from the stress-energy tensor, the theory may contain further chiral
fields, such as currents, subject to an appropriate boundary condition; 
e.g., for a conserved current with $j_L=\frac12(j_0+j_1)=j_L(t+x)$, 
$j_R=\frac12(j_0-j_1)=j_L(t-x)$, 
the vanishing of the charge flow across the boundary gives
\begin{equation} 
j_{1}(t,x=0) = 0 \qquad\Leftrightarrow \qquad j_L=j_R\equiv j, 
\end{equation}
and 
\begin{equation} 
j_0(t,x) = j(t+x) + j(t-x), \qquad j_1(t,x) = j(t+x) - j(t-x). 
\end{equation}

It is crucial to contrast the bilocal form (1.2), (1.4) of the
chiral fields in boundary CFT with the situation in 2D Minkowski space
CFT, where, e.g., the stress-energy tensor has the chiral decomposition
\begin{equation} 
T_{00}(t,x) = T(t+x)\otimes\Eins + \Eins\otimes T(t-x), \quad
T_{01}(t,x) = T(t+x)\otimes\Eins - \Eins\otimes T(t-x), 
\end{equation}
where $T_L=T\otimes\Eins$ and $T_R=\Eins\otimes T$ are two {\em independent}
(left and right) chiral fields. A boundary CFT contains only one chiral 
algebra with an appropriate identification between left and right movers. 
Consequently, the representation space is a direct sum of representations 
of the chiral algebra, rather than of tensor products of representations of 
two chiral algebras. This ought to be ascribed to the fact that the
imposing of boundary conditions and the ensuing breakdown of symmetry
have so drastic consequences on the ground state fluctuations (Casimir
effect) that states respecting the boundary conditions cannot be
realized in the Hilbert space of states without boundary conditions,
see, e.g., \cite{Ja}.

Let us point out, however, that {\em locally} the two situations with
$T_L=T_R=T$ and with  $T_L=T\otimes\Eins$ and $T_R=\Eins\otimes T$ independent, 
are {\em algebraically
  indistinguishable}: for instance, in the latter case the commutator
$[T_L(t_1,x_1)\pm T_R(t_1,x_1), T_L(t_2,x_2)\pm T_R(t_2,x_2)]$
involves only $\delta$-function contributions at $t_1+x_1=t_2+x_2$ and at
$t_1-x_1=t_2-x_2$, while the commutator $[T(t_1,x_1)\pm
T(t_1,x_1), T(t_2,x_2)\pm T(t_2,x_2)]$ has additional contributions at 
$t_1+x_1=t_2-x_2$ and at $t_1-x_1=t_2+x_2$. But within a wedge region 
$M_+\supset W: x>\vert t\vert$ ($\Leftrightarrow$ $t-x<0<t+x$), the latter
contributions are ineffective. The same holds for any time translate
of $W$.

A slightly stronger version of this algebraic indistinguishability is
the following:  It has been shown that the chiral
stress-energy tensor satisfies the {\em split property}: namely for 
every pair of intervals $J < I$ which do not touch (thus allowing to
smooth out the UV singularities), there exists a state $\varphi$ in
the vacuum Hilbert space $\HH_0$ of $T$ (depending on $I$ and $J$; in
particular not the vacuum state) which has no correlations among
$T(u_1)$ and $T(u_2)$ when $u_1\in I$ and $u_2\in J$. In other words,
$\varphi$ factorizes on products of $T(u_i)$ with $u_i\in I \cup J$
according to    
\begin{equation} 
\varphi\big(\prod_k T(u_k)\big)= \varphi\big(\prod_{i:u_i\in I}T(u_i)\big)
\cdot \varphi\big(\prod_{j:u_j\in J}T(u_j)\big).
\end{equation} 
This implies that, for every double-cone $O$ not touching the boundary
(hence $t-x$ and $t+x$ belong to non-touching intervals as before),
there is a state $\varphi$ such that products of $T_{\mu\nu}(t,x)$
given by (1.2) in the boundary CFT with $(t,x)\in O$ have the same
expectation values in the state $\varphi$ as the same products of 
$T_{\mu\nu}(t,x)$ given by (1.5) in the 2D Minkowski space CFT have in
the state $\varphi\otimes\varphi$. This property exhibits the 
local ``decoupling'' of left and right chiral components.  Exactly as
the split property fails when the intervals $I$ and $J$ touch, the
decoupling of left- and right-movers breaks down at the boundary in BCFT.

We shall assume the split property for {\em all} chiral fields of a
boundary CFT. This property is known to be related to phase space
properties of the CFT (existence of $\Tr\exp-\beta L_0$)
\cite{BAL,ALR}, and it has been established for large classes of
chiral models (\cite{Xu1} and references therein). 

Let us now turn to local fields in boundary CFT which do not decompose
in the manner of (1.2) or (1.4). These non-chiral fields have to
satisfy local commutativity with the chiral fields and with each
other, and transform covariantly under the conformal (M\"obius) group
generated by the chiral stress-energy tensor $T$. 
The starting point in the present article will be the crucial
observation that  
\begin{center}
non-chiral local fields in BCFT arise from non-local chiral fields 
\end{center}
by an algebraic construction (explained in detail in Sect.\ 2). This
construction gives also rise to a model-independent explanation
(Sect.\ 5) for an observation due to Cardy \cite{C} concerning the
structure of correlation functions. Cardy has shown that $n$-point
functions of primary local fields in boundary CFT satisfy the same
differential equations in the $2n$ variables $t_i\pm x_i$ as chiral
$2n$-point conformal blocks of an associated two-dimensional Minkowski
space CFT, and are therefore particular combinations of the
latter. E.g., the 4-point function of the order parameter in the
critical Ising model in the full plane factorizes as   
\begin{eqnarray} 
\langle\Omega\;,\;\sigma(t_1,x_1)\,\sigma(t_2,x_2)\,
\sigma(t_3,x_3)\,\sigma(t_4,x_4)\;\Omega\rangle = \qquad\qquad \nonumber \\
= F(t_1+x_1,\dots,t_4+x_4)\cdot F(t_1-x_1,\dots,t_4-x_4 ) + 
\nonumber \\  +
G(t_1+x_1,\dots,t_4+x_4)\cdot G(t_1-x_1,\dots,t_4-x_4 ) 
\end{eqnarray}
(where the chiral 4-point conformal blocks $F$ and $G$ correspond to
intermediate states in the vacuum sector and in the ``energy'' sector,
respectively), whereas both 
\begin{equation} 
\langle\Omega\;,\;\phi_0(t_1,x_1)\,\phi_0(t_2,x_2)\;\Omega\rangle \propto
F(t_1+x_1,t_1-x_1,t_2+x_2,t_2-x_2)
\end{equation} 
and
\begin{equation} 
\langle\Omega\;,\;\phi_1(t_1,x_1)\,\phi_1(t_2,x_2)\;\Omega\rangle \propto
G(t_1+x_1,t_1-x_1,t_2+x_2,t_2-x_2)
\end{equation} 
are 2-point functions of local fields on the half-plane
$M_+$. Expressed in terms of exchange fields \cite{RS} (``generalized
chiral creation and annihilation operators''), we have the operator
factorization  
\begin{equation} 
\sigma(t,x) = a(t+x)\otimes a(t-x) + b(t+x)\otimes b(t-x) + \hbox{h.c.}
\end{equation}
on the Hilbert space $\left[\HH_0\otimes\HH_0\right]\oplus
\left[\HH_{\frac1{16}}\otimes \HH_{\frac1{16}}\right]\oplus
\left[\HH_{\frac12}\otimes\HH_{\frac12}\right]$,
where $\HH_h$ are the three sectors of the stress-energy tensor with
$c=\frac12$, the exchange fields $a:\HH_0\to\HH_{\frac1{16}}$ and
$b:\HH_{\frac1{16}}\to\HH_{\frac12}$ and their adjoints interpolate
among the three sectors of the chiral stress-energy tensor, and $F=\langle
a^*aa^*a\rangle$, $G=\langle a^*b^*ba\rangle$. In contrast, (1.8) and
(1.9) are two-point functions of local fields on the half-plane, 
given by  
\begin{equation} 
\phi_0(t,x) \propto a^*(t+x)\,a(t-x) 
\end{equation}
defined on $\HH_0$, and 
\begin{equation} 
\phi_1(t,x) \propto b(t+x)\,a(t-x) + a^*(t+x)\,b^*(t-x)
\end{equation}
defined on $\HH_0\oplus\HH_{\frac12}$. The local commutativity at
space-like distance of the combinations (1.11), (1.12) can be
directly checked in terms of the exchange (braid group) commutation
relations among $a$, $b$ and their adjoints \cite{RS}.\footnote{More
  precisely, while any linear combination $\phi_1$ of the two terms in
  (1.12) satisfies local commutativity with itself, it does so with
  $\phi_1^*$ only if $\phi_1$ is a multiple of a hermitean field. 
  ((1.12) {\em is} hermitean up to a phase due to the exchange
  comutation relations $b(t-x)a(t+x)=\omega b(t+x)a(t-x)$ and 
  $a^*(t-x)b^*(t+x)=\omega a^*(t+x)b^*(t-x)$ with
  $\omega=\exp-i\frac38\pi$ \cite{RS}.) On the other hand, any two
  hermitean combinations differ only by a unitary Klein
  transformation; thus up to a global phase and unitary 
  similarity, the combination (1.12) is unique as a local quantum field.}  
In this calculation, the specific ordering
$t_1-x_1<t_2-x_2<t_2+x_2<t_1+x_1$ (or $1\leftrightarrow 2$) is
crucial. In particular, the combinations $\phi_0$ and $\phi_1$ given
by (1.11), (1.12) on the {\em entire} plane would fail to be 
{\em local} fields. 

We learn from this explicit example that the local fields in boundary CFT 
carry a bi-localized product of charges of the chiral algebra, rather than a 
tensor product of left and right charges, as in Minkowski space CFT. 
Moreover, they interpolate in very specific ways among the charged
sectors of the chiral algebra, and these structures determine the
scaling behavior of the fields as $x\to 0$. E.g., the field $\phi_0$
has a singular behavior $\propto x^{-\frac2{16}}$ as $x\to 0$, while
the field $\phi_1$ vanishes $\propto x^{\frac12-\frac2{16}}$ at the
boundary.\footnote{In the general case, one argues as follows. As
  $x\to 0$, the variables $t+x$ and $t-x$ coalesce. Thus, the scaling
  bahvior is controlled by the operator product expansion, and
  depends on the particular fusion channel selected by the
  bi-localization formula.} Thus, we also see that the choice of a boundary 
condition is related to the bi-localized charge structure of the local fields. 
We shall investigate the origin of this charge structure in the
general case.  

For these purposes, we look at boundary CFT from the algebraic point of view 
\cite{H}. The algebraic point of view emphasizes the representation theoretic 
features of a QFT, especially charges and their composition \cite{DHR},
rather than kinematical features such as analytic properties of correlation 
functions. The DHR theory of superselection sectors \cite{DHR} asserts that
all information about charges (superselection sectors), their composition 
(``fusion'', operator product expansions), and their interchange 
(``statistics'', commutation relations) is encoded in a braided C* tensor 
category (the DHR category for short), in terms of local observable 
quantities. This theory has been developed further into a powerful tool, 
useful for explicit computational purposes especially in the chiral setting. 

E.g., the classification of local and non-local extensions of a given 
local QFT has been cast into a problem of classification of 
{\it Q-systems} (\cite{LR}, see Sect.\ 4 and App.\ A) within the DHR
category. Q-systems are an efficient tool to control the algebraic
consistency of commutation relations, operator product expansions, and
charge conjugation of primary and descendant fields at one stroke. 
Under the natural assumption of ``complete rationality'' (Sect.\ 2),
the classification of irreducible chiral extensions in CFT has thus been
shown to be a finite-dimensional problem with finitely many solutions 
(see Sect.\ 3.2). In the case $c<1$, a complete classification has been
obtained along these lines \cite{KL}. 

Furthermore, the existence of exchange fields as in (1.10)--(1.12)
with numerical braid group commutation relations and their operator
product expansion could be established from general principles
in the algebraic approach \cite{FRS1,FRS2}.

Rather than the local fields, say $\phi(t,x)$, the prime objects in
the algebraic approach to QFT are the von Neumann algebras of local
observables generated by the fields smeared with localized test 
functions, say 
\begin{equation} 
A(O):=\{\phi(f),\phi(f)^*: \hbox{supp}\,f \subset O\}'' 
\end{equation}
for open space-time regions $O$. The properties of the assignment
$O\mapsto A(O)$ (the {\em net of local algebras}) are axiomatized such that 
their generation by fields as in (1.13) becomes in fact obsolete and needs 
not be assumed at all. 

In our case, the chiral fields $T(u)$, ($j(u)$, \dots) generate a chiral 
net of local von Neumann algebras
\begin{equation} 
I\mapsto A(I), \qquad I=(a,b)\subset\RR 
\end{equation}
on the vacuum Hilbert space $\HH_0$. In fact, $A$ extends to a net
over the intervals of the circle (embedding $\RR$ into
$S^1$ by means of a Cayley transformation). 

The chiral fields of a boundary CFT generate a net 
\begin{equation} 
O\mapsto A_+(O).
\end{equation}
According to the prescription (1.13), $A_+(O)$ is generated by chiral
fields smeared in the variable $t+x$ over the interval $I$ and in the
variable $t-x$ over the interval $J$, where $O=I\times J$, $I>J$, is
an open double-cone in $M_+$. The bi-localized structure (1.2), (1.4) 
etc., translates into the form of the local algebras (cf.\ Fig.\ 2)
\begin{equation} 
A_+(O) = A(I) \vee A(J) \qquad (O=I\times J,\;\;I>J).
\end{equation}

\begin{minipage}{115mm}\hskip40mm \epsfig{file=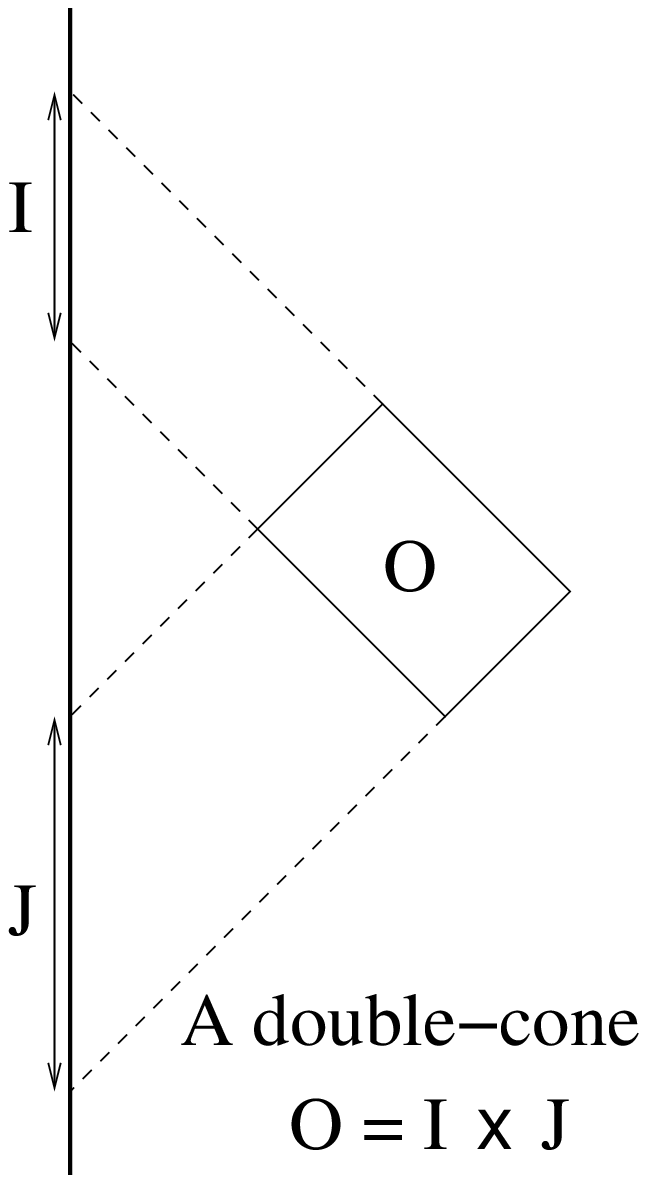,width=2.9cm} 

\small {\bf Fig.\ 2:} A double-cone in the half space $M_+$. An
observable in $A_+(O)$ is generated by chiral observables
localized in $I$ and $J$.
\end{minipage}
\vskip3mm

The (searched for) non-chiral local fields of the boundary CFT 
will generate a net of local algebras in their vacuum representation
\begin{equation} 
O\mapsto B_+(O),
\end{equation}
with $B_+(O)$ containing $A_+(O)$ and consequently commuting with 
$A_+(\hat O)$ as $\hat O$ is space-like from $O$. But while $A_+$ is
defined on the vacuum representation space $\HH_0$ of the chiral CFT
$A$, the boundary CFT $B_+$ will in general be defined on a larger
Hilbert space $\HH_B\supset \HH_0$, i.e., one has
\begin{equation} 
\pi(A_+(O))\subset B_+(O).
\end{equation}
Since we assume covariance and positive energy throughout, the 
representation $\pi$ of $A$ on $\HH_B$ is a positive-energy
representation, containing the vacuum representation, with irreducible
decomposition  
$\pi\simeq\bigoplus_s n^s\cdot\pi_s$, $n^0=1$. 

We take the structure (1.14)--(1.18) as the characteristic structure
of algebraic boundary conformal QFT, irrespective whether the nets are 
generated as in (1.13) by any specific set of generating local fields. 
Our main results will be the following (for any fixed chiral CFT $A$), 
referring to the body of the article for more detailed qualifications
of the statements: 

In Sect.\ 2, we provide several general results about the structure of
local boundary CFT's $B_+$. Every maximal local boundary CFT $B_+$ can
be recovered from its ``restriction to the boundary''. The latter is 
some (possibly non-local) chiral extension $I\mapsto B(I)$ of the
chiral CFT $I\mapsto A(I)$ (Prop.\ 2.9), defined on the same Hilbert
space $\HH_B$ as $B_+$. Structural features of the latter (reviewed
and developed in Sect.\ 3, where Tomita's Modular Theory \cite{T}
plays a crucial role) are exploited to infer structural features of
boundary CFT. We shall refer to the (re)construction of the boundary
CFT from a (non-local) chiral theory as {\em (boundary) induction}. 

These results show that the classification of (non-local) chiral
extensions of a local chiral theory (e.g., in terms of Q-systems)
at the same time provides a classification of boundary CFT's.

On the other hand, every (non-local) chiral extension $B$ of $A$ determines 
a local CFT $B^\alpha_2$ on two-dimensional Minkowski space-time with left 
and right chiral observables $A\otimes A$ (henceforth referred to as the 
{\em $\alpha$-induction construction}). The Hilbert space of $B^\alpha_2$ 
carries the representation $\pi_2\simeq\bigoplus_{\sigma\tau}
Z_{[\sigma][\tau]} \cdot \pi_\sigma\otimes\pi_{\bar \tau}$ of $A\otimes A$,
where the matrix $Z$ with indices in the set of irreducible sectors of
$A$ is a modular invariant determined by the chiral extension
$B$ \cite[Cor.\ 1.6]{CTPS}.

In Sect.\ 4, we discuss the relation between these two constructions
of 2D nets (boundary induction for the half-space vs.\
$\alpha$-induction for Minkowski space). Indeed, the local inclusions
$\pi(A_+(O))\subset B_+(O)$ and $\pi_2(A(I)\otimes A(J)) \subset
B^\alpha_2(O)$ are algebraically isomorphic (Thm.\ 4.1). In this
sense, the boundary CFT constitutes a representation of the local
degrees of freedom of the Minkowski space theory $B_2^\alpha$ which is
consistent with the chiral boundary condition (1.1) and its
generalizations such as (1.3). 

But the representation spaces of $B_+$ and of $B^\alpha_2$ are 
very different, one being a direct sum of sectors of $A$, the other 
being a direct sum of tensor products of sectors. Therefore inspite of
the {\em algebraic} isomorphism, the bi-localized charge structure of the 
local fields on the half-space must be structurally different from the 
tensor product charge structure of the local fields in the plane, as is
clearly exemplified by (1.11) or (1.12) vs.\ (1.10). 

We derive an explicit formula for the local charged fields (Prop.\ 5.1) 
exhibiting their bi-localized charge structure in terms of non-local
chiral exchange operators. The charge of a field $\phi(t,x)$ is a
product (not a {\em tensor} product) of two chiral charges localized
at $t+x$ and $t-x$, respectively. This structure, and as a consequence
the behavior of the  charged fields and their correlations close to
the boundary, is determined by the non-local chiral extension $B$,
i.e., the choice of $B$ ``determines the boundary conditions''.

In this sense, the natural reasoning where one would impose the boundary
conditions first, and then attempt to construct local fields subject
to these conditions, is inverted. This avoids the problem with
the usual strategy, that a consistent set of boundary conditions must be
chosen in the first place, while it is not a priori clear what
``consistent'' would mean. 
As our analysis demonstrates implicitly, the algebraic constraints on
the local fields to be constructed are highly involved: they consist
in (a) the Q-system describing the algebraic structure of the inclusion
$A_+(O)\subset B_+(O)$, and (b) the representation of this algebraic
structure on a Hilbert space $\HH_B$. From these data which most sensitively
depend on the DHR structure of the underlying chiral net $A$, the
boundary conditions emerge, so that it is very unlikely that it should
be possible to ``guess'' the consistent sets of boundary conditions
without further specific insight. For this reason, we consider the
present top-down strategy 
\begin{center}
chiral extension $\to$ boundary condition
\end{center}
much more effective, since it is completely under control
in the algebraic framework. 

In Sect.\ 6 we show that along with a given (non-local) chiral 
extension $B$, there is a whole family of non-local chiral extensions
$B_a$, all associated with the same Minkowski space theory $B^\alpha_2$, 
and hence a family of boundary CFT nets $B_{a,+}$, which are all locally
isomorphic, but whose local fields exhibit different bi-localized
charge structures and satisfy different boundary conditions, in the
sense just explained. The multiplicities of the Hilbert spaces 
$\HH_a\equiv \HH_{B_a}= \bigoplus_s n^s_a\cdot\HH_s$ 
are the diagonal elements of a ``nimrep'' (non-negative
integer matrix representation) of the fusion rules of $A$:
\begin{equation}
n^s\cdot n^t = \sum_u N^{st}_u \;n^u \qquad\hbox{with}\qquad n^s_{aa}
= n^s_{a}. 
\end{equation}
We include in Sect.\ 7 some preliminary remarks on the relation to the
modular structure of partition functions and boundary states. 

The structural analysis pursued in this article generalizes closely
related previous analyses in complementary approaches. In the context of
critical phenomena in Statistical Mechanics, Cardy has already
discussed \cite{C} the case $B=A$ (in our terminology), leading to the
set of boundary conditions being labelled by the sectors of $A$. 
The same situation was investigated by Felder, Fr\"ohlich, Fuchs and
Schweigert \cite{FFFS} from the perspective of three-dimensional 
topological field theory. Fuchs, Runkel and Schweigert \cite{FRS}
proceeded to construct the coefficients of all $2n$-point conformal
blocks as in (1.8), (1.9) in a combinatorial manner, where a condition very
similar to our eq.\ (5.12) was crucial to ensure locality. 
Behrend {\em et al.}\ \cite{BPPZ} have concentrated on graph theoretic
aspects of the pertinent fusion algebras, and to A-D-E classification
aspects in the case of $SU(2)$ current algebras, see also \cite{Z} for
a review. Fuchs and Schweigert \cite{FS1} have studied the generalization in
which (in our terminology) $A$ is a subtheory (not necessary of
orbifold type) of a chiral theory $B$ which is itself local. They also
emphasized the role of $\alpha$-induction. This case is known to give
rise to block diagonal modular invariant matrices $Z_{st}$ \cite{BEK1}, and
to this case also applies the result in \cite{K}. The same authors
\cite{FS2} have further developed the purely categorial aspects
characteristic of boundary CFT, no longer referring to the underlying
physical postulates. In fact, these structures fit most naturally in
the general setting of tensor categories as exposed, e.g., in
\cite{FS2,KO,MM2}. 

In comparison to such a considerable gain of mathematical generality
(where quantum physics remains hardly visible), the motivation and
ambition of our work is more limited. On the other hand, we study and
explain specific representation theoretic issues which in the other
frameworks are not or even cannot be addressed. For these issues,
operator algebraic methods are most powerful. 

We emphasize that in our approach the prominent principle is Locality. 
In other approaches \cite{Z}, inspired by Statistical Mechanics or String
Theory rather than Quantum Field Theory, Modular Invariance of the
partition function is taken instead as a first principle, required in
order to guarantee that the theory can be consistently defined on
arbitrary Riemann surfaces. It is well known, however that -- although
closely related to each other -- these principles cannot be precisely
mapped onto each other \cite{MI}.  

In fact, we do not assume diffeomorphism invariance but only
M\"obius invariance. Assuming diffeomorphism invariance (i.e., the
algebraic implementation of localized diffeomorphisms by suitable
chiral observables), would allow some stronger results.
E.g., (for an explanation of the notions, see the beginning of the next
section), it was shown in \cite{LX} that strong additivity would be
automatic in a split net of finite $\mu$-index, and that the
$\mu$-index coincides with the dimension of the DHR category. 
Concerning boundary CFT, one could infer that the index of the
inclusion of the chiral observables in the BCFT observables associated
with a double-cone, does not depend on the double-cone, in spite of
the fact that the M\"obius group does not act transitively on the
double-cones in $M_+$.

\section{Algebraic boundary conformal QFT}
\setcounter{equation}{0} 

We work with a fixed chiral conformal net $I\mapsto A(I)$ over the
intervals of the real axis \cite{GL}, e.g., a Virasoro net with $c<1$
or a non-abelian current algebra (affine Kac-Moody) chiral net. In this 
article, $A$ is assumed to be {\em completely rational} \cite{KLM}. This
condition combines {\em rationality} (finitely many superselection sectors,
each with finite statistics \cite{DHR,FRS1}), {\em strong additivity}
(``irrelevance of points for smearing'', i.e., the algebras of two
adjacent intervals $(a,b)$ and $(b,c)$ generate the algebra of the
full interval $(a,c)$; this property is equivalent to Haag duality of the
chiral theory on the real line), and the {\em split property} (statistical
independence of local algebras $A(I)$ and $A(J)$ when $I$ and $J$ are
finitely separated, and as a consequence $A(I)\vee A(J)$ is isomorphic
to $A(I)\otimes A(J)$, cf.\ the discussion around (1.6); this property
is guaranteed, e.g., if $\exp-\beta L_0$ is a trace class operator for
all $\beta$ in the vacuum representation \cite{BAL,ALR}). Most of the
common models of chiral CFT are completely rational \cite{L,Xu1},
but abelian current algebras as well as stress tensors with $c\geq 1$
without further fields are excluded by the assumption of rationality. 

Completely rational chiral theories enjoy very interesting properties
concerning the structure of their superselection sectors. E.g., the
DHR statistics is non-degenerate (besides the vacuum sector, no sector
has trivial monodromy with every other sector) \cite[Cor.\ 37]{KLM},
and thus gives rise to a unitary representation of the modular group
$SL(2,\ZZ)$ in terms of the statistics \cite[Cor.\ 5.2]{FRS2}, turning
the DHR category into a {\em modular category} \cite{Tu}. Moreover, in
completely rational theories, the dimension of the DHR category (the
sum of the squares of the dimension of all irreducible superselection
sectors), equals the ``$\mu$-index'' (the von Neumann subfactor index
of the inclusion $A(E)\subset A(E')'$ where $E$ is the union of two
disconnected intervals and $E'$ its complement on the circle
\cite[Thm.\ 33]{KLM}).   

\vskip3mm
{\bf 2.1. Geometric preliminaries on the half-space $M_+$.}
\vskip2mm
Before turning to QFT on the half-space $M_+\equiv\{(t,x)\in\RR^2:x>0\}$, 
let us mention some elementary geometric properties of this space:

(i) A double-cone {\em within} the half-space $M_+$ is an open region of
the form $O=I\times J\equiv\{(t,x): t+x\in I,\; t-x\in J\}$ whose
closure is contained in $M_+$ (cf.\ Fig.\ 3). Let $L\subset\RR$ be a
bounded open interval, and $J<K<I$ the three subintervals (ordered as
indicated) obtained by removing two points from $L$ (cf.\ Figs.\ 2 and
5). There is a bijection between the configurations of four intervals
$I,J,K,L$ obtained this way and the double-cones $O$ within $M_+$,
such that $O=I\times J$. By default, $I,J,K,L$ and $O$ will always
refer to such a configuration. Only if necessary, we shall write
$I_O,J_O,K_O,L_O$ in order to indicate this convenient parametrization
of double-cones within $M_+$. 

(ii) A left wedge is a region of the form $W_L=\{(t,x):\vert t-t_0\vert
< x_0-x\}$ for some $(t_0,x_0)\in M_+$; it is ``spanned'' by the
interval $I=(t_0-x_0,t_0+x_0)$. A right wedge is a region of the form
$W_R=\{(t,x):\vert t-t_0\vert < x-x_0\}$ for some $(t_0,x_0)\in M_+$.
The causal complement of a left wedge is a right wedge, and vice
versa (cf.\ Fig.\ 3). The causal complement of a double-cone is the
union of a left wedge $W_L=O_<$ and a right wedge $W_R=O_>$. A
double-cone $\hat O$ belongs to the left causal complement $O_<$ of
$O$ ($\hat O < O$) iff $L_{\hat O}\subset K_O$, and to the right
causal complement $O_>$ ($\hat O > O$) iff $L_O\subset K_{\hat O}$. 

Locality of a net $B_+$ on $M_+$ means that $B_+(\hat O)$
commutes with $B_+(O)$ in both cases.

\begin{minipage}{115mm}\hskip10mm \epsfig{file=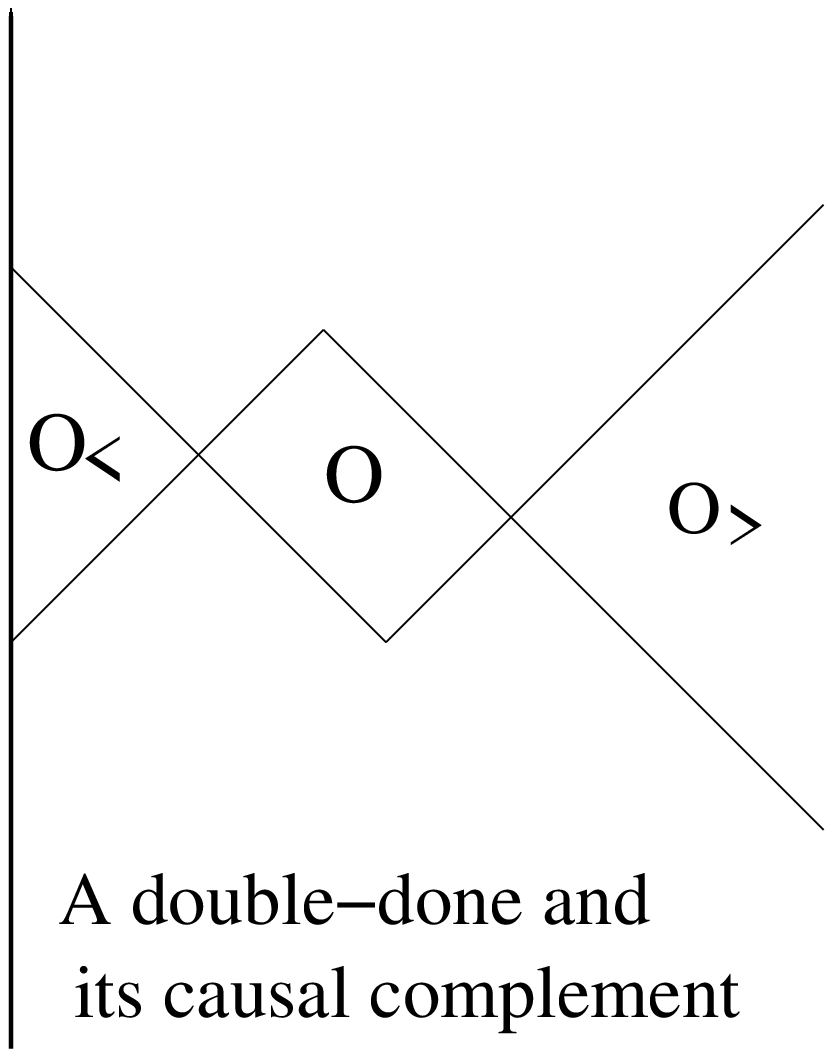,width=3.2cm} 
\hskip20mm \epsfig{file=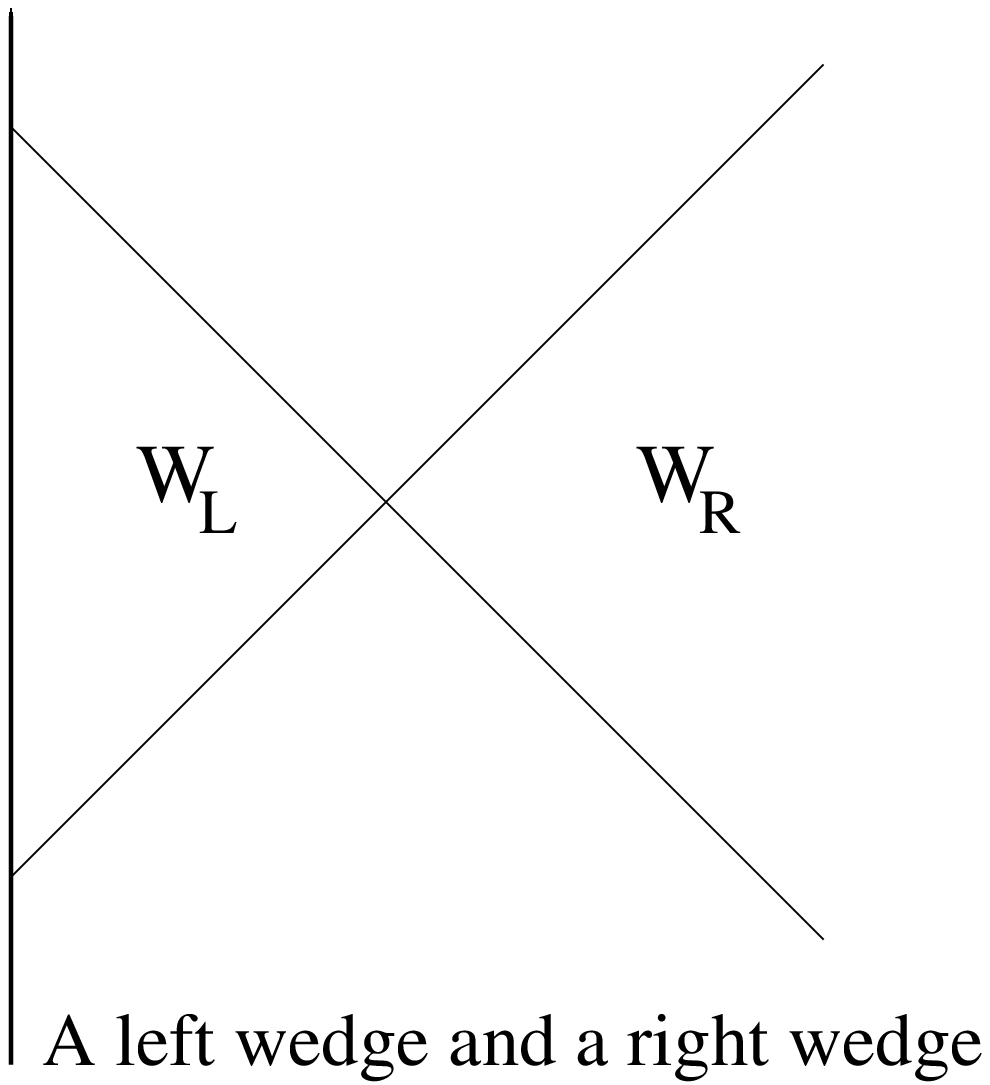,width=4.0cm}  

\small {\bf Fig.\ 3:} Double-cone and wedge regions in $M_+$ and their
causal complements.
\end{minipage}
\vskip3mm

(iii) The covering of the M\"obius group $G:=\widetilde{PSL}(2,\RR)$,
acting on the universal covering of the compactification $S^1$ of
$\RR$, induces an action on a certain covering of $M_+\subset
\RR\times \RR$. The subgroups of translations and of dilations act on
$\RR$, and the induced actions are the time translations and the
dilations of $M_+$, respectively.  
\vskip3mm 
{\bf 2.2. Local algebras in boundary CFT.}
\vskip2mm
 {\bf 2.1 Definition:} A given chiral net $A$ defines two different local nets over the open
double-cones within $M_+$, namely the {\em trivial boundary CFT}
\begin{equation} 
O\mapsto A_+(O):= A(I)\vee A(J) 
\end{equation}
and  its {\em dual}
\begin{equation} 
O\mapsto A\dual_+(O):= A(L)\cap A(K)'. 
\end{equation}
As emphasized by the notation, $A\dual_+$ is the {\it dual net} associated 
with $A_+$:
\begin{equation} 
A\dual_+(O) := A_+(O')' 
\end{equation} 
where $A_+(O'):= \bigvee_{\hat O\subset O'}A_+(\hat O)\equiv 
A_+(O_<)\vee A_+(O_>)$ is the algebra
generated by all observables of $A_+$ localized in double-cones at
space-like separation from $O$. 

{\it Remarks:} 1. Both nets $A_+$ and $A\dual_+$ are represented on 
the same Hilbert space $\HH_0$, the vacuum Hilbert space of $A$. 
 The observables of the trivial BCFT are bilocal expressions 
in the chiral observables, as described in the Introduction.

2. The dual net is local because, if $O_1$ and $O_2$ are space-like 
separated within $M_+$, then $L_2\subset K_1$ (or $1\leftrightarrow 2$), 
hence $A\dual_+(O_1) \subset A(K_1)'$ and $A\dual_+(O_2) \subset A(L_2)$ 
commute. It follows that $A\dual_+$ is its own dual net ({\it Haag-duality}). 

3. The inclusion
\begin{equation} 
A_+(O) \subset A\dual_+(O) 
\end{equation}
is the ``two-interval subfactor'' extensively discussed in \cite{KLM}. 
Apart from $A(I)$ and $A(J)$, the algebra $A\dual_+(O)$ contains all
unitary ``charge transporters'' $u: \rho^I\to\rho^J$ where $\rho^I$,
$\rho^J$ are (equivalent) DHR endomorphisms of $A$ localized in $I$
and $J$, respectively,\footnote{For details on DHR theory in the
  chiral setting, see \cite{FRS1,FRS2}. The notation $t:\rho\to\sigma$
  means the intertwining property $t\rho(a)=\sigma(a)t$ for all $a\in
  A$. We shall also write $t\in \Hom(\rho,\sigma)$.}  
and these elements generate $A\dual_+(O)$. The algebraic isomorphism
class of the two-interval subfactor (2.4) does not depend on the pair
of intervals, and thus on $O$. 

4. We observe that
\begin{equation} 
\bigvee_{O:\;L_O\subset L} A_+(O) = \bigvee_{O:\;L_O\subset L}
A\dual_+(O) = A(L), 
\end{equation}
because the intervals $I_O$ and $J_O$, as $O$ varies as specified, cover
all of $L$. 

The  trivial BCFT $A_+$ and  its dual $A\dual_+$ are special cases 
\footnote{The latter is sometimes called ``the Cardy case'' in the literature
  \cite{FFFS}.}) of boundary conformal  quantum field theories in the
sense of the following definition.  

{\bf 2.2 Definition:} A {\em boundary  CFT (BCFT)} associated 
with $A$ is a local, isotonous net $O\mapsto B_+(O)$ over the
double-cones within the half-space $M_+$, represented on a
Hilbert space $\HH_B$ such that

(i) there is a unitary representation $\UU$ of the covering of the
M\"obius group $G=\widetilde{PSL}(2,\RR)$ with positive generator for
the subgroup of translations, such that  
\begin{equation} 
\UU(g)B_+(O)\UU(g)^* = B_+(gO) 
\end{equation}
whenever the conformal transformation $g\in G$ takes the double-cone
$O=I_O\times J_O$ within $M_+$ into another double-cone $gO:= gI_O
\times gJ_O$ within $M_+$ \footnote{$G$ being a covering group, this
  means more precisely the following: $g$ is represented by a path
  $g_t\in PSL(2,\RR)$ connecting the identity with $p(g)\in PSL(2,\RR)$, 
  such that $g_tO$ lies within $M_+$ for all $t$.} (i.e., in
particular for all translations and dilations), with a unique
invariant vector $\Omega\in\HH_B$ (the vacuum vector). 

(ii)  There is a representation $\pi$ of $A$ on $\HH_B$ such that
$B_+(O)$ contains $\pi(A_+(O))$, and 
\begin{equation} 
\UU(g)\pi(A_+(O))\UU(g)^* = \pi(A_+(gO))
\end{equation}
whenever $O$ and $gO$ are double-cones within $M_+$. 

(iii) {\it ``Joint irreducibility'':} For each double-cone $O$, the von Neumann algebra $B_+(O)\vee
\pi(A_+)''$ is irreducible on $\HH_B$, i.e., equals $\BB(\HH_B)$. Here, $\pi(A_+)$ is the C* algebra generated by all
double-cone algebras $\pi(A_+(O))$, and $\pi(A_+)''$ is its weak
closure, i.e., the von Neumann algebra generated by all interval
algebras $\pi(A(I))$. 

{\it Comments:} 1. By Remark 4 following Def.\ 2.1, the covariance condition in (ii)
is equivalent to  
\begin{equation} 
\UU(g)\pi(A(I))\UU(g)^* = \pi(A(gI))
\end{equation}
whenever $I$ and $gI$ are intervals in $\RR$. As a consequence, $\pi$
extends to a positive-energy representation of the chiral net $A$ on
the circle.

2. Joint irreducibility (iii) implies irreducibility
of the net $B_+$ on $\HH_B$ with $\Omega$ the unique $\UU$-invariant
vector, cf.\ \cite{GL}. On the other hand, $\Omega$ being the unique
$\UU$-invariant vector and cyclic for $B_+$ implies the irreducibility
of $B_+$ by Prop.\ 3.3 below (choosing $U$ the subgroup of time
translations). The covariance and spectrum condition (i) implies that
the vacuum vector is in fact cyclic and separating for every local
algebra $B_+(O)$ (Reeh-Schlieder property). 

3. Joint irreducibility also implies that the local
inclusions $\pi(A_+(O))\subset B_+(O)$ have trivial relative
commutant.

4. Joint irreducibility is automatic if the
representation $\UU(g)$ belongs to $\pi(A_+)''$, e.g., if the
stress-energy tensor of the BCFT coincides with that of the chiral
theory $A$, see e.g., \cite{CO,AFK}. 

5. In general, a BCFT net $B_+$ does {\it not} contain the
dual net $\pi(A\dual_+)$, nor is it relatively local with respect to
$\pi(A\dual_+)$ (see Prop.\ 2.7 for a characterization of this case;
clearly, the former property would imply the latter). 

{\bf 2.3 Definition:} The {\em dual net} $O\mapsto B\dual_+(O)$ of a 
boundary  CFT $O\mapsto B_+(O)$ is defined by  
\begin{equation} 
B\dual_+(O) := B_+(O')' \equiv B_+(O_<)' \cap B_+(O_>)'.
\end{equation} 
Here, $B_+(O_<)$, $B_+(O_>)$ are the von Neumann algebras generated 
by all $B_+(\hat O)$ as $\hat O$ belongs to the left or right causal 
complement of $O$, respectively. By locality of $B_+$, $B\dual_+(O)$ 
contains $B_+(O)$.

We shall see below (Prop.\ 2.10, using {\it Modular Theory} \cite{T})
that $B_+$ is in fact {\em wedge dual}, i.e., the algebra of a right
wedge $W_R$ is the commutant of the algebra of the corresponding left
wedge $W_R=W_L'$, and vice versa. This means that (2.9) may be
rewritten as
\begin{equation} 
B\dual_+(O) := B_+(O_>') \cap B_+(O_<').
\end{equation} 
In particular, the dual net $B\dual_+$ is again local, and
consequently it is its own dual (i.e., it is Haag dual). The notation is 
consistent since $A\dual_+$ is indeed the dual net of $A_+$.  
\vskip3mm
{\bf 2.3. The non-local chiral net associated with a BCFT.}
\vskip2mm
We now turn to the description of a BCFT in terms of a {\em chiral} net,
which is in general {\em non-local}.

{\bf 2.4 Definition:} A boundary  CFT $O\mapsto
B_+(O)$ generates a chiral net $I \mapsto B\gen(I)$ (the associated
{\em boundary net}) on $\HH_B$, by 
\begin{equation} 
B\gen(I) := \bigvee_{O\subset W_L} B_+(O) \equiv B_+(W_L)
\end{equation}
where $W_L$ is the left wedge spanned by $I$ (cf.\ Fig.\ 4).

By the above Remark 4 following Def.\ 2.1, this definition associates
the original net $A$ with both $A_+$ and $A\dual_+$.

\begin{minipage}{115mm}\hskip40mm \epsfig{file=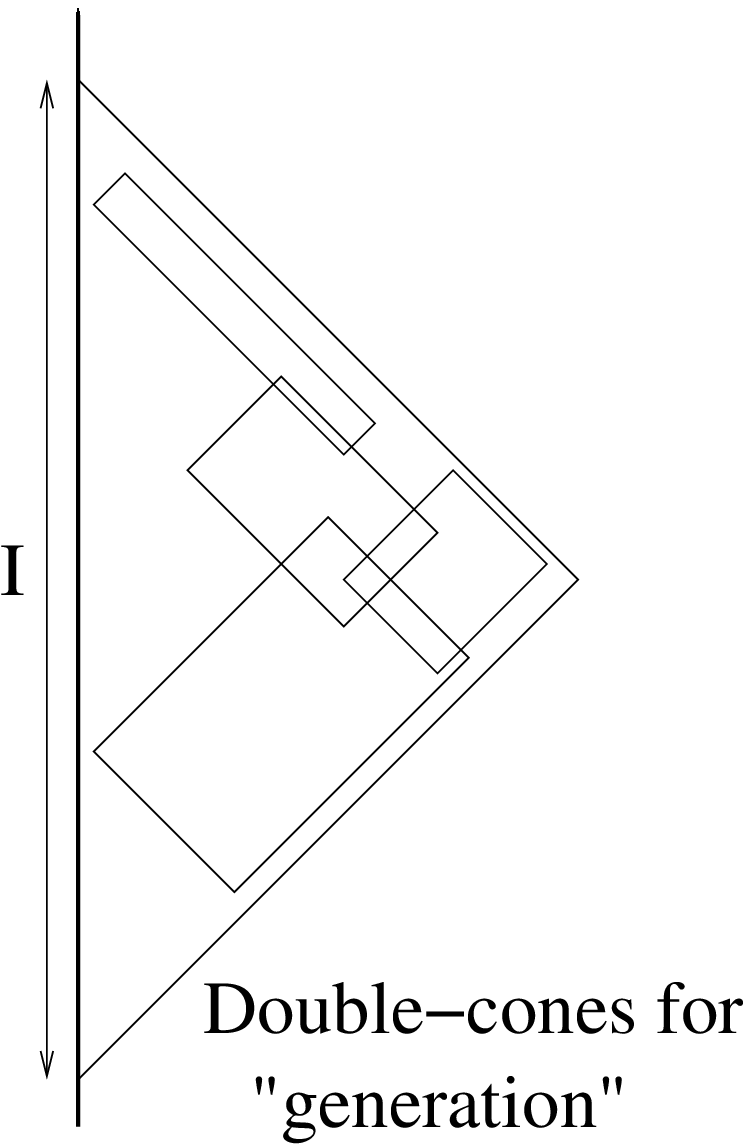,width=3.3cm} 

\small {\bf Fig.\ 4:} The observables of the associated chiral boundary
net localized in $I$ are generated by BCFT observables localized in
double-cones $O\subset W_L$.
\end{minipage}
\vskip3mm
{\bf 2.5 Proposition:} (i) The boundary net $B\gen$ generated from
$B_+$ is isotonous, and it is covariant:
\begin{equation} 
\UU(g)B\gen(I)\UU(g)^* = B\gen(gI) 
\end{equation}
whenever $I\subset \RR$, $gI \subset\RR$ \footnote{in the same sense
  as explained in the footnote to Def.\ 2.2(ii)}. It acts
irreducibly on $\HH_B$. $B\gen$ extends $\pi(A)$ and is relatively
local with respect to $\pi(A)$: 
\begin{equation} 
\pi(A(I)) \subset B\gen(I) \subset \pi(A(I'))'. 
\end{equation}

(ii) There is a consistent family of vacuum-preserving conditional
expectations ${\cal E}^I: B\gen(I)\to A(I)$.

(iii) The local subfactors $\pi(A(I))\subset B\gen(I)$ are irreducible and
have finite index. The index is independent of $I$. 

In general, the boundary net $B\gen$ is a {\em non-local} chiral 
net. For if $I_1$ and $I_2$ are disjoint, then the double-cones
contributing to the definition (2.11) of the corresponding algebras
$B\gen(I_1)$ and $B\gen(I_2)$ are pairwise time-like separated, and
observables of $B_+$ need not satisfy time-like commutativity. 
Non-local chiral nets have been studied before, e.g., in \cite{ALR}.
We shall review and extend their general structure theory in Sect.\ 3. 
In the remainder of the present section, we shall freely use these
results.  

{\it Proof of Prop.\ 2.5:} (i) Isotony, covariance, the extension
property and relative locality are elementary. Irreducibility of the
net $B\gen$ follows from irreducibility of $B_+$. 

(ii) The statement will be proven in the next section (Prop.\ 3.5(i)).

(iii) Irreducibility of the local subfactors $A(I)\subset B\gen(I)$
follows from joint irreducibility. Namely, the von Neumann algebra
$\pi(A(I))\vee B\gen(I)'$ contains $\pi(A(I))\vee \pi(A(I'))$ hence
$\pi(A_+)$ by strong additivity, and $B_+(O)$ for some $O$ with
$I\subset K_O$, hence it equals $\BB(\HH_B)$. Thus 
$\pi(A(I))'\cap B\gen(I)=\CC\cdot\Eins$.   

Irreducibility implies finiteness of the index because $A$
is completely rational, by the same argument as in \cite[Prop.\ 2.3]{KL} 
(i.e., each irreducible subsector $\rho$ can arise in $\pi$ with
multiplicity bounded by $d(\rho)$).  
Its independence of the interval follows as in \cite[Cor.\ 4.2]{LR}.  
\QED 

Prop.\ 2.5 means that the extension $\pi(A)\subset B\gen$ defines 
what was called a {\em quantum field theoretical net of subfactors} in 
\cite{LR}. We shall use here rather the terminology {\it chiral extension}.
Conditional expectations having the abstract properties of an average
(``non-commutative integration''), the existence of a consistent
family of vacuum-preserving conditional expectations of $B(I)$ to
$A(I)$ was viewed in \cite{LR} as a ``generalized symmetry which is
unbroken in the vacuum state''.

It should be emphasized that a consistent family of (vacuum-preserving) 
conditional expectations cannot be expected in general for the
double-cone algebras $A_+(O)\subset B_+(O)$, because the modular
automorphism group of $B_+(O)$ acts non-geometrically and therefore
does not preserve the subalgebra $A_+(O)$. In the case of $B_+=A\dual_+$, 
the failure can be seen directly: here the cyclic subspace of $A_+$ 
coincides with the full Hilbert space, and the corresponding projection
is the unit operator. On the other hand, the unique conditional
expectation of $A\dual_+(O)$ to $A_+(O)$ \cite{KLM} does not preserve
the vacuum. Likewise, there cannot be a {\em global} conditional
expectation, because, e.g., ${\cal E}^{\hat O}$ is trivial on
$A\dual_+(O)$ whenever $I_{\hat O}$ or $J_{\hat O}$ contains $L_O$
because $A\dual_+(O) \subset A(L_O) \subset A_+(\hat O)$, while
${\cal E}^O$ is non-trivial on the same algebra.

In this sense, the (generalized) symmetry allows to
determine the subalgebras $\pi(A_+(W)) \subset B_+(W)$ associated with
wedges as fixpoint algebras, but the same does not hold for the
subalgebras $\pi(A_+(O)) \subset B_+(O)$ associated with double-cones. 
(This is completely analogous to compact symmetry groups acting on field
algebras associated with connected and with disconnected regions in
four dimensions \cite{DHR}). As a consequence, the techniques and
results of \cite{LR} do not apply directly to algebraic boundary CFT,
considered as the net of subfactors $I\mapsto A_+(O)\subset B_+(O)$. 
Instead, as a consequence of Prop.\ 2.5, these techniques {\em do} 
apply to the associated boundary extension $A(I)\subset B\gen(I)$, and
we shall elaborate in Sect.\ 4 and 5 how they {\em indirectly} provide
the desired insight into the structure of the net of algebras on the
half-space and its representations.      

A central result of \cite{LR} is the
following generation property (for further explanations, see Sect.\ 4
and App.\ A).  

{\bf 2.6 Corollary \cite{LR}:} 
For each interval $I$, the ``dual canonical'' endomorphism of $A(I)$
associated with the local subfactor $A(I)\subset B\gen(I)$ extends to
a DHR endomorphism $\theta^I$ of $A$ localized in $I$. The algebra $B\gen(I)$
is generated by its subalgebra $\pi(A(I))$ and a ``canonical''
isometry $v^I\in B\gen(I)$ which is an intertwiner for $\theta^I$,
i.e., one has $\pi(\theta^I(a))v^I = v^I\pi(a)$ for all $a\in A$. 

This property can be used to obtain

{\bf 2.7 Proposition:} If $B_+$ is relatively local with respect to 
$\pi(A\dual_+)$, then $B\gen=A$, and $B_+$ lies between $A_+$ and $A\dual_+$.

{\it Proof:} Let $O=I\times J$, $J<I$, and $K$ and $L$ as described in
the beginning of Sect.\ 2.1. Assume that $\pi(A\dual_+(O))$ commutes
with $B_+(\hat O)$ whenever $\hat O$ belongs to the left causal
complement of $O$, i.e., whenever $L_{\hat O} \subset K$. Then
$\pi(A\dual_+(O))$ commutes with $B\gen(K)$. Every unitary charge
transporter $u: \rho^I\to\rho^J$ belongs to $A\dual_+(O)$, thus
$\pi(u)$ commutes with $B\gen(K)$. By Cor.\ 2.6, $v^K\in B\gen(K)$
satisfies $v^K\pi(u) = \pi(\theta^K(u))v^K$ while by locality
$v^K\pi(u) = \pi(u)v^K$, hence $\pi(\theta^K(u))v^K = \pi(u)v^K$. As
an equation in $B\gen(L)$, this implies \cite{LR} $\theta(u)=u$. By
\cite{FRS2}, this implies that the sectors $[\rho]$ and $[\theta]$
have trivial monodromy, and as $[\rho]$ was arbitrary, $[\theta]$ has
trivial monodromy with every DHR sector. But by \cite{KLM}, the
braiding of the completely rational net $A$ is non-degenerate, hence
$[\theta]$ must be trivial. This in turn implies $B\gen=A$ \cite{LR}.  

The last statement will follow from Prop.\ 2.9(ii), 
 according to which $B\gen=A$ implies $B\dual_+ = A\dual_+$.
\QED

By the definition of the boundary net $B$ and locality of $B_+$, we
obviously have $B_+(O) \subset B\gen(L) \cap B\gen(K)'$. This
suggests the following definition of a local boundary CFT {\em induced} 
by a given (possibly non-local) chiral net:

{\bf 2.8 Definition:} If $I\mapsto B(I)$ is an irreducible chiral
extension of $I\mapsto A(I)$ (possibly non-local, but
relatively local with respect to $A$), then the {\em induced net} is
defined by (cf.\ Fig.\ 5) 
\begin{equation} 
O\mapsto B\ind_+(O) := B(L) \cap B(K)'.
\end{equation}

\begin{minipage}{115mm}\hskip40mm \epsfig{file=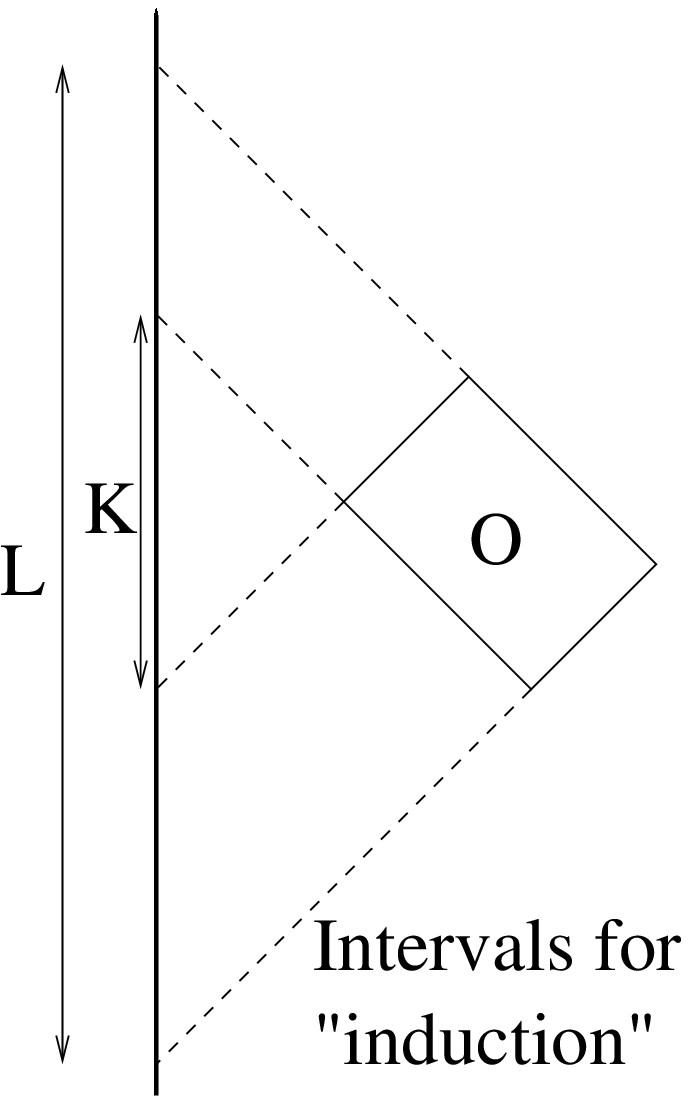,width=28mm} 

\small {\bf Fig.\ 5:} The observables of the induced BCFT localized in
$O$ belong to $B(L)$ and commute with $B(K)$.
\end{minipage}
\vskip3mm

Let us discuss to which extent Def.\ 2.8 is the converse of Def.\ 2.4,
i.e., to which extent a boundary CFT can be reconstructed
from its boundary net:

{\bf 2.9 Proposition:} (i) The induced net (2.14)
is a boundary  CFT associated with $A$, defined on the
Hilbert space of $B$. E.g., in the special case $B=A$, the induced net
is the dual net $A\dual_+$.

(ii) If $B$ is a chiral extension of $A$, then the boundary net
$(B\ind_+)\gen$ generated by the induced net $B\ind_+$ is again $B$.  
Conversely, if $B_+$ is a boundary  CFT, then its boundary
net $B\gen$ induces the dual net $B\dual_+$ associated with $B_+$,
i.e., $(B\gen)\ind_+=B\dual_+$. In other words, we have: gen $\circ$
ind $=$ id, ind $\circ$ gen $=$ dual, implying dual $\circ$ ind $=$
ind and dual $\circ$ dual $=$ dual.  

(iii) Every induced net $B\ind_+$ is self-dual (Haag dual). 

{\it Proof:} (i) $B\ind_+$ contains $\pi(A_+)$ and is local by
definition. The covariance properties (2.6) and (2.7) follow from
covariance of the chiral net $B$. Joint irreducibility is automatic 
because $\pi(A(I))\subset B(I)$ has finite index by virtue of
irreducibility of $\pi(A(I))\subset B(I)$ and complete
rationality \cite{KL}, which implies that $\UU(g)$ belongs to the von
Neumann algebra $\pi(A)''$ generated by all $\pi(A(I))$, cf.\ Comment
4 after Def.\ 2.2. 

(ii) $(B\ind_+)\gen(L)$ is generated by the algebras $B(L) \cap B(K)'$ as $K$ 
varies within $L$, so its commutant is the intersection of the algebras 
$B(L)'\vee B(K)$ as $K$ varies. For any fixed $K_0\subset L$, by the split 
property for the net $B$ (Prop.\ 3.6), $B(L)'\vee B(K)$ is naturally 
isomorphic to $B(L)'\otimes B(K_0)$. Now, as $K$ varies within $K_0$, the 
intersection $\bigcap_{K\subset K_0} B(K)$ is trivial  
(this follows from ``triviality at a point'', Prop.\ 3.2(ii)), hence 
$\bigcap_{K\subset K_0} B(L)'\otimes B(K)=B(L)'\otimes \CC\Eins$. 
It follows that $\bigcap_{K\subset K_0} B(L)'\vee B(K)$ equals 
$B(L)'\vee \CC\Eins = B(L)'$, and 
$(B\ind_+)\gen(L) = \bigvee_K B(L)\cap B(K)'= B(L)$. 

Conversely, if $B_+$ is given, then by definition 
\begin{equation}
B\gen(K) = B_+(O_<) 
\end{equation} 
since $O_<$ is the left wedge spanned by $K$. We shall show next
(Prop.\ 2.10) that boundary CFT nets satisfy {\em wedge duality}.
Hence, because the right wedge $O_>$ is the causal complement of the
left wedge spanned by $L$, we have also 
\begin{equation}
B_+(O_>) = B\gen(L)'.
\end{equation}
This implies $B\dual_+(O) = B\gen(L) \cap B\gen(K)' = (B\gen)\ind_+(O)$. 

(iii) is obvious from (ii). \QED

As the examples (Def.\ 2.1) of the trivial BCFT and its dual show,
there is no bijection between boundary CFT's and their boundary nets;
but Prop.\ 2.9(ii) means that there is a bijection between 
{\em Haag-dual} boundary CFT's and their boundary nets.  Yet, the
non-Haag-dual boundary CFT's being subtheories of the Haag-dual ones,
the previous results show that a classification of boundary CFT's
essentially reduces to a classification of (non-local) {\em chiral}
extensions.

The following facts (some of which anticipate results from Sect.\ 5)
provide some non-trivial examples for the results in this section:

The chiral theory $A$ of the stress-energy tensor with $c=\frac12$ has
one non-trivial chiral extension $B$, the CAR algebra of a chiral real
Fermi field $\psi$ on $\HH_B=\HH_0\oplus\HH_{\frac12}$. The local
algebras $A\dual_+(O)$ of the dual net are generated by $A_+(O)$,
the operators $(\psi(f)\psi(g))\rest_{\HH_0}$ with ${\rm supp}\;f\subset I$, 
${\rm supp}\;g \subset J$, and the field $\phi_0$ (eq.\ (1.11)) smeared
within $O$.  (In Sect.\ 5 it will become clear that this characterization is 
equivalent with the one given in Remark 3 following Def.\ 2.1.) 
On the other hand, the local algebras $B\ind_+(O)$ are
generated by $\pi(A_+(O))$, $\psi(f)\psi(g)$ and the field $\phi_1$
(eq.\ (1.12)). The subtheory on $\HH_0$ with local algebras
$B_+(O)=(CAR(I)\otimes CAR(J))^{\rm even}$ generated by $\pi(A_+)$ and
$\psi(f)\psi(g)$ is not Haag dual. While the field $\phi_0$ may well
be ``lifted'' to $\HH_{\frac12}$, it is impossible to do so such that
it locally commutes also with $\phi_1$.  

In Sect.\ 4, we shall show how to ``compute'' the
intersection $B\ind_+(O) = B(L) \cap B(K)'$ from the non-local chiral
extension $B$ of $A$ in the general case (in terms of the DHR category
of the local chiral net $A$), and to obtain algebraic invariants for
the inclusions $\pi(A_+(O)) \subset B\ind_+(O)$ from the chiral
subfactors $\pi(A(I)) \subset B(I)$. 
\vskip3mm
{\bf 2.4. General results: duality and split property for wedges}
\vskip2mm
We proceed with several general structure results about boundary CFT's.

In the sequel, for $W$ a left wedge, $\Delta_W^{is}, J_W$ are the
modular data associated with the von Neumann algebra $B_+(W)$ and the
vacuum state \cite[Chap.\ VI, Thm.\ 1.19]{T}, and $\Lambda_W(s)$ is
the one-parameter subgroup of the (covered) M\"obius group $G$
preserving $W$, defined as follows: Let $\lambda(s):u\mapsto
\exp(-s)u$ be the scale transformations. Then
$\Lambda_W(s):=g\lambda(s)g\inv$ where $g\in PSL(2,\RR)$ maps
$(0,\infty)$ to the interval $I$ of the real line which spans $W$.  

We denote by $r_W$ the inversion which maps $I$ to its complement on
the circle, i.e., $r_W=hrh\inv$ where $r$ is the inversion $u\mapsto
1/u$, and $h\in G$ maps $(-1,1)$ to $I$. We denote by the same symbol
the (densely defined) transformations of Minkowski space  or the
half-space $M_+$, induced by acting simultaneously on $t+x$ and $t-x$. 

Finally, $\Gamma$ is the group generated by $r$ and $G$. Then we prove

{\bf 2.10 Proposition:} (i) Every boundary CFT $B_+$ satisfies 
{\em wedge duality}
\begin{equation} 
B_+(W') = B_+(W)'
\end{equation}
where $W$ is a left wedge and $W'$ its causal complement.

(ii) Every boundary CFT $B_+$ has the {\it Bisognano-Wichmann property}:
\begin{equation} 
\Delta_W^{is}=\UU(\Lambda_W(-2\pi s))
\end{equation}
for every left wedge $W$. There exists an (anti-)unitary
representation $\tilde \UU$ of the group $\Gamma$ on $\HH_B$
extending the representation $\UU$ of $G$ such that 
\begin{equation} 
J_W=\UU(r_W).
\end{equation} 
In particular, $J_W\UU(g)J_W=\UU(r_Wgr_W)$ ($g\in \Gamma$) and
\begin{equation} 
J_W B\dual_+(O) J_W = B\dual_+(r_WO)
\end{equation} 
whenever $O$ and $r_WO$ are double-cones within $M_+$.

{\em Proof:} The first part (2.18) of (ii) is proven in Prop.\ 3.5(ii)
where $B(I)\equiv B_+(W)$, $A(I)\equiv A_+(W)$, using 2.5(iii).  

Turning to (i), we note that wedge duality holds for $A_+$, because it
is equivalent to Haag duality on the real line for $A$, which is in
turn equivalent to strong additivity.  

Let $W$ be a left wedge and $W'$ its causal complement. 
Consider the inclusions
\begin{equation} 
\pi(A_+(W'))\subset B_+(W')\subset B_+(W)'.
\end{equation}
The subfactor  
\begin{equation} 
\pi(A_+(W'))=J_W\pi(A_+(W))J_W\subset J_WB_+(W)J_W = B_+(W)'
\end{equation}
is irreducible with finite index, because $\pi(A_+(W))=\pi(A(I))\subset
B\gen(I)=B_+(W)$ is irreducible with finite index. 
Clearly, $B_+(W)'$ is globally stable under $\Ad_{\UU(\Lambda_W(s))}$,
and the same is true for $A_+(W')$ by strong additivity of $A$. 
Due to the rigidity of intermediate subfactors in subfactors with
finite index \cite{L}, the intermediate algebra $B_+(W')$ in (2.21)
must also be globally stable under $\Ad_{\UU(\Lambda_W(s))}$.

Thus, $B_+(W')$ is a von Neumann subalgebra of $B_+(W)'$ cyclic on the
vacuum, which, thanks to (2.18), is in addition invariant under the
modular automorphisms of $B_+(W)'$. By modular theory \cite[Chap.\ IX,
Thm.\ 4.2]{T}, this algebra must coincide with $B_+(W)'$.  This proves
wedge duality.   

Turning to the second part of (ii), we infer from Prop.\ 3.2(iv) that 
$J_W\UU(g)J_W=\UU(r_Wgr_W)$ ($g\in G$), thus (2.19) defines a
representation. Furthermore, $\Ad_{J_W}$ acts covariantly on interval
algebras: $\Ad_{J_W}B\gen(K)=B\gen(r_WK)$, hence on left wedge algebras: 
 $\Ad_{J_W}B_+(W_L)=B_+(r_WW_L)$, hence on right wedges by wedge
duality:  $\Ad_{J_W}B_+(W_R)=B_+(r_WW_R)$. As $B\dual_+(O)$ is defined as
an intersection of wedge algebras, it follows that $\Ad_{J_W}$ acts
covariantly on the dual net as stated in (2.20). \QED

E.g., if $W$ is spanned by the interval $(-1,1)$, then $r_W:u\mapsto 1/u$ 
induces the ray inversion $(t,x)\mapsto(\frac t{t^2-x^2},-\frac x{t^2-x^2})$. 
This map maps only the region $x>\vert t\vert$ of $M_+$ into $M_+$. Thus,
(2.20) makes sense only for double-cones $I\times J$ such that
$I\subset\RR_+$ and $J\subset \RR_-$. On other double-cones, $J_W$
acts non-geometrically.

{\em Remark:} If $B_+$ is not Haag-dual, one may consistently define
$\bar B_+(O)$ by $J_WB_+(r_WO)J_W$ (choosing $W$ such that $r_WO$
belongs to $M_+$). This defines another BCFT intermediate between
$A_+(O)$ and $B\dual_+(O)$.

{\bf 2.11 Proposition:} Every boundary CFT $B_+$ satisfies the split
property for wedges. That is, if $O$ is a double-cone within $M_+$ and
$W_L=O_<$ and $W_R=O_>$ the associated pair of left and right wedges,
then the inclusion $B\ind_+(W_L)\subset B\ind_+(W_R)'$ is split, or
equivalently $B\ind_+(W_L)\vee B\ind_+(W_R)$ is naturally isomorphic
to the tensor product $B\ind_+(W_L)\otimes B\ind_+(W_R)$.
In particular, this implies the split property for double-cones $O_1$,
$O_2$ whenever $O_1\subset O_<$ and $O_2\subset O_>$, i.e.,  
$B\ind_+(O_1)\vee B\ind_+(O_2)$ is naturally isomorphic to the
tensor product $B\ind_+(O_1)\otimes B\ind_+(O_2)$.

{\it Proof:} The inclusion $A\subset B\gen$ has finite index 
 (Prop.\ 2.5(iii)). Thus $B\gen$ is split by
Prop.\ 3.6, i.e., the inclusion $B\gen(K)\subset B\gen(L)$ is
split. Now by definition, $B_+(W_L)= B\gen(K)$, and by wedge duality
(Prop.\ 2.10(i)), $B_+(W_R)'=B\gen(L)$. This proves the claim. \QED  

{\bf 2.12 Proposition:} Let $B$ be a chiral extension of $A$, and
$B\ind_+$ the induced BCFT net. Then 

(i) The index of $\pi(A_+(O))\subset B\ind_+(O)$ equals the
$\mu$-index $\mu_A$ of $A$ (i.e., the index of the two-interval subfactor 
$\mu_A=[A(L')\cap A(K): A(I)\vee A(J)]= [A\dual_+(O):A_+(O)]$ which
coincides with the dimension of the DHR category of $A$ \cite{KLM}; in
particular, it is independent of $O$).  This index is thus the same
for each chiral extension.  

(ii) The induced net $B\ind_+$ satisfies strong additivity. 

{\it Proof:} (i) Let $\lambda=[B(I):\pi(A(I))]$ denote the index of the chiral
extension. $\lambda$ is independent of $I$ and finite 
 (Prop.\ 2.5(iii)). We want to compute the index of
$\pi(A(I)\vee A(J))\subset B(K)'\cap B(L)$, which equals the index of
the commutant $\pi(A(I))'\cap \pi(A(J))' \supset B(K) \vee B(L)'$. 

Using the notation 
\begin{equation}
N_1\stackrel{\alpha}\subset N_2
\end{equation}
to indicate that a subfactor $N_1\subset N_2$ has index $[N_2:N_1]=\alpha$,
we shall prove the indices displayed in the square of inclusions
\begin{equation}
\begin{array}{ccc}
\pi(A(I))'\cap \pi(A(J))'&\stackrel{?}\supset & B(K) \vee B(L)' \\[2mm]
{\scriptstyle \lambda^2\mu_A}\; \cup&&\cup\;{\scriptstyle \lambda} \\
\pi(A(K))\vee \pi(A(L'))&\stackrel\lambda\subset& \pi(A(K))\vee B(L)'
\end{array}
\end{equation}
from which the index of  the subfactor in 
the top row follows to be $\mu_A$ by the
multiplicativity of the index, as claimed in the statement.

The inclusion in the left 
column is the two-interval subfactor of the chiral net $A$ in the
representation $\pi$ on $\HH_B$, whose index has been computed in
\cite[Lemma 42]{KLM} as follows: $\pi$ is unitarily equivalent to a
DHR endomorphism $\theta$ of $A$ in its vacuum representation
\cite{LR} (see also Sect.\ 4), where $\theta$ has dimension
$d(\theta)=\lambda$; thus we may as well consider the subfactor
$\theta(A(K)\vee A(L'))\subset \theta(A(I)\vee A(J))'$ on
$\HH_0$. Choosing $\theta$ to be localized in $K$, we get $\theta(A(K))
\vee A(L') \subset A(K)\vee A(L') \subset (A(I)\vee A(J))'$. The
former inclusion has index $d(\theta)^2=\lambda^2$, and the latter is
the two-interval subfactor of index $\mu_A$. Thus, the index in the
left column equals $\lambda^2\mu_A$. 

The indices in the right column and bottom row $[B(K):\pi(A(K))]$ and
$[B(L)':\pi(A(L'))]$, respectively, by the split property for $B$
(Prop.\ 3.6). The former is $\lambda$ by definition, while the latter
equals $\lambda$ because in $\pi(A(L'))\subset B(L)'\subset
\pi(A(L))'$ the second inclusion has index $\lambda$ by definition,
while the total inclusion is the one-interval subfactor of $A$ in the
representation $\pi$ which has dimension $\lambda^2$ \cite[Lemma 42]{KLM}.

Thus the index in the top row equals $\mu_A$.

The statement (ii) now follows exactly as in \cite[Lemma 23]{L}. \QED  
\vskip3mm
{\bf 2.5. Superselection structure of boundary CFT.}
\vskip2mm
In the remainder of this section, we discuss DHR sectors (= superselection
sectors in the sense of \cite{DHR}) for Haag dual boundary CFTs. 
M\"uger has shown \cite{MM} that in Minkowski space-time,
the split property for wedges implies the absence of nontrivial
sectors. We obtain here a similar result on the half-space.

A DHR sector of a boundary CFT $B_+$ is defined as an equivalence class
of positive-energy representations $\pi$ subject to the selection
criterium that $\pi$ cannot be distinguished from the (defining)
vacuum representation by measurements within the causal complement of
any double-cone $O$ within $M_+$. Assuming Haag duality for $B_+$, by
standard arguments \cite{DHR} one finds that superselection sectors
can be represented by localized and transportable endomorphisms (DHR
endomorphisms) $\rho$ of the net $B_+$. This means that for any given
double-cone $O$, $\rho$ can be chosen within its unitary equivalence
class to act like the identity map on the algebra of the causal
complement $B_+(O')= B_+(O_<)\vee B_+(O_>)$, and the unitary charge
transporters which intertwine equivalent such endomorphisms localized
in different regions belong to $B_+$. \footnote{Unlike double-cones in  
Minkowski space-time, any given pair of double-cones $O_1$, $O_2$
within $M_+$ is not always contained in another double-cone within
$M_+$. However, given $O_1$ and $O_2$, one can choose an auxiliary
$O_3$ such that $O_1\cup O_3$ and $O_2\cup O_3$ are each
contained in some double-cone within $M_+$. The charge transporter
from $O_1$ to $O_2$ may then be obtained as a composition of two
charge transporters from $O_1$ to $O_3$ and from $O_3$ to $O_2$.}

It is obvious that every DHR endomorphism of $B_+$ defines a localized
and transportable endomorphism (in the obvious sense) of the boundary
net $B\gen$; but the converse is not true. E.g., the DHR
endomorphisms $\rho$ of the chiral net $A$ (which is the boundary net 
generated by $A\dual_+$) localized in, say, the interval $K$ act
non-trivially on the charge transporters ``across'' $K$ \cite{FRS2}. 
However, such charge transporters do belong to $A\dual_+(O_>)$, so
$\rho$ is not localized as an endomorphism of $A\dual_+$. In fact, the
following result shows that the dual net $A\dual_+$, and in fact any
Haag dual boundary CFT, does not possess any nontrivial DHR sectors at all.

Let $E = O_1 \cup O_2$ be the union of two causally disjoint
double-cones within $M_+$ which do not touch (i.e., whose closures are
disjoint); we may assume that $O_1$ belongs to the left causal
complement of $O_2$, $O_1 < O_2$. Then the causal complement $E'$ of
$E$ is the union of the left wedge $W_L=O_1{}_<$, the right wedge
$W_R=O_2{}_>$ and a double-cone $O=O_1{}_>\cap O_2{}_<$. We consider the
inclusion $B_+(E) \subset B_+(E')'$, 
where $B_+(E)$ and $B_+(E')$ are defined by additivity. 

{\bf 2.14 Proposition:} If $B_+\supset A_+$ is a boundary CFT net,
then the index $\mu_{B_+}$ of the inclusion 
\begin{equation}
B_+(E) \subset B_+(E')'
\end{equation}
is independent of $E$ and equals 
\begin{equation}
\mu_{B_+}= [B\dual_+:B_+]^3
=\left(\frac{\mu_A}{[B_+:\pi(A_+)]}\right)^3
\end{equation} 
where the indices of the extensions $[B\dual_+:B_+]:=
[B\dual_+(O):B_+(O)]$ and $[B_+:\pi(A_+)]:= [B_+(O):\pi(A_+(O))]$ are
independent of $O$. In particular, $\mu_{A_+}=\mu_A^3$. 

{\bf 2.15 Corollary:} (i) When $B_+$ is Haag dual, then $\mu_{B_+}=1$, and
$B_+$ satisfies Haag duality also for disconnected regions of the form
$E=O_1\cup O_2$ as above (i.e., (2.25) is an equality).  

(ii) A Haag dual boundary CFT net $B_+$ has no nontrivial DHR sectors.

(iii) When $B_+$ is not Haag dual, then $B\dual_+$ is a field net for
$B_+$ in the sense of \cite{DHR}, i.e., for every sector of $B_+$
represented by a DHR endomorphism $\rho$, there is a nontrivial
operator in $B\dual_+$ which intertwines $\rho$ with the identity.

{\it Proof of the Proposition:} The independence of the indices on the
various regions (of a given topology) follows as in \cite{KLM,LR}.

We denote by $B$ the induced boundary net, and write $[B:A]=:\lambda$
and $[B_+:\pi(A_+)]=:\lambda_+$. We shall show that 
\begin{equation}
\begin{array}{ccc}
B_{+}(E)&\stackrel{\mu_{B_{+}}}{\subset}& B_{+}(E')'\\[2mm]
\stackrel{\lambda_+^2}{}\,\cup&&\cap\, \stackrel{\lambda^2\lambda_+}{} \\
\pi(A_+(E))&\stackrel{\lambda^2\mu_{A}^{3}}{\subset}& \pi(A_+(E'))'
\end{array}
\end{equation}
which implies 
\begin{equation}
\mu_{B_{+}}=\left(\frac{\mu_A}{\lambda_+}\right)^3
\end{equation} 
by multiplicativity of the index:

Bottom row of (2.27): $A_+(E)= A(J_2)\vee A(J_1)\vee A(I_1)\vee A(I_2)$
is a four-interval algebra of the chiral net $A$, and so is $A_+(E')$
by strong additivity of $A$. Thus we have the four-interval subfactor
in the representation $\pi$ whose index is computed, as in Prop.\
2.12, with the help of \cite[Lemma 42]{KLM} to
be $d(\theta)^2\mu_A^3=\lambda^2\mu_A^3$. 

Left column of (2.27) $[B_{+}(E):\pi(A_+(E))]=\mu^2_A$: 
We have
\begin{equation}
B_{+}(E)= B_{+}(O_1)\vee B_{+}(O_2)\supset 
\pi(A_+(E))= \pi(A(O_1))\vee \pi(A(O_2))
\end{equation}
so, by the split property for $B_+$ (Prop.\ 2.11), $[B_{+}(E):\pi(A_+(E))]= 
[B_+(O_1):\pi(A_+(O_1))]\cdot [B_{+}(O_2):\pi(A_+(O_2))]$ where each factor 
equals $\lambda_+$. 

Right column of (2.27) $[B_{+}(E'):\pi(A_+(E'))]=\mu_A$: 
The computation is analogous to the previous one, but here $E'=W_L\cup
O \cup W_R$ has 3 connected components. The double-cone contributes a
factor $\lambda_+$ as before, while the two wedges contribute a factor 
$[B_+(W):\pi(A_+(W))]=[B(I):\pi(A(I))]=\lambda$ each. 

This proves the various indices in (2.27) and hence the formula
(2.28). By Prop.\ 2.12,
$\mu_A=[B\dual_+(O):\pi(A_+(O))]=[B\dual_+(O):B_+(O)][B_+(O):\pi(A_+(O))]$  
gives $[B\dual_+(O):B_+(O)]=\mu_A/\lambda_+$. This proves (2.26).
\QED

{\it Proof of the Corollary:} 
The statement (i) is obvious from the proposition. The proof for the
absence of non-trivial sectors in the Haag dual case is exactly as
(iii) $\Rightarrow$ (ii) in \cite[Cor.\ 32]{KLM}%
\footnote{There is some unfortunate misnumbering of the implications
  proven in \cite[Cor.\ 32]{KLM}. (i) $\Rightarrow$ (ii) should read
  (iii) $\Rightarrow$ (ii), and (ii) $\Rightarrow$ (iii) should read
  (i) $\Rightarrow$ (iii), while (iii) $\Rightarrow$ (i) is trivial.},
using Haag duality of $B_+$ for disconnected regions of the form $E$:
If $u:\rho_1\to\rho_2$ is a unitary intertwiner from $\rho_1$
localized in $O_1$ to $\rho_2$ localized in $O_2$, then $u$ belongs to 
$B_+(E')'=B_+(E)$. Thanks to the split property for $B_+$ (Prop.\ 2.11), 
there is a conditional expectation ${\cal E}: B_+(E)=B_+(O_1)\vee
B_+(O_2)\to B_+(O_1)$ such that ${\cal E}(u)\neq 0$. This is a
nontrivial local intertwiner from $\rho_1\rest{B_+(O_1)}$ to
$\id\rest{B_+(O_1)}$, hence a global intertwiner from $\rho_1$ to
$\id$ thanks to strong additivity for $B_+$ (Prop.\ 2.12). Thus every
sector contains the identity sector, which implies the claim. 

Similarly, when $B_+$ is not Haag dual, and $\rho_i$ are a pair of
equivalent DHR endomorphisms of $B_+$ as before, then the charge
transporter $u:\rho_1\to\rho_2$ belongs to $B\dual_+(E)$, and 
${\cal E}(u):\rho_1\to\id$ belongs to $B\dual_+(O_1)$, as asserted. \QED 

{\em Remark:} To prevent misconceptions of the statement (ii) of 
Cor.\ 2.15, it should be pointed out that a Haag-dual BCFT can well
have non-trivial positive energy-representations; e.g., every
positive-energy representation of $A$ defines a positive-energy
representation of $A\dual_+$. But these representations are not localized
in double-cones as required for DHR representations. 

\section{Non-local chiral CFT}
\setcounter{equation}{0}

This section reviews and generalizes known structural theorems about
non-local chiral CFT, and also contains several new results. While
the section is logically independent of BCFT, its results bear
important implications for BCFT. They are freely used in other sections. 
\vskip3mm
{\bf 3.1 General structure: Covariance and modular symmetry.}
\vskip2mm
In \cite{ALR}, the covariant transformation law for a non-local chiral CFT 
\begin{equation} 
\UU(g)B(I)\UU(g)^* = B(gI) \qquad (g\in G)
\end{equation}
was assumed to hold {\em globally}, i.e.\ for every interval of the
circle and without restriction on $g\in G$. It was shown that this
implies the rotation of the circle by $4\pi$ to be represented by
$\UU(4\pi)=\Eins$, hence the conformal Hamiltonian $L_0$ has half-integer
spectrum and the net $B$ is (at least weakly) graded local.  

In our setting, this restriction is too narrow. Depending on the
spectrum of $L_0$ on $\HH_B$, (3.1) holds for the induced net only
locally as indicated in (2.12), admitting ``more non-local'' induced
boundary CFT's than graded local ones. In order to generalize the
analysis in \cite{ALR}, we first note

{\bf 3.1 Lemma:} Let $I\to B(I)$ be a net of von Neumann algebras
defined on the intervals $I\subset\RR$, and $\UU$ a representation of
$G=\widetilde{PSL}(2,\RR)$ on the same Hilbert space such that (3.1)
holds whenever $I$ and $gI$ belong to $\RR$. Then, identifying $\RR$
with $S^1\setminus\{-1\}$ by means of a Cayley transformation, $B$
extends to a net defined on the intervals of the (universal) covering
$\SS$ of $S^1$ for which (3.1) holds globally. 

{\em Sketch of the Proof:} Use (3.1) as a definition of the algebra
$B(gI)$ on the right-hand side whenever the conditions on $I$ and $g$
are {\em not} met, i.e., whenever $gI$ belongs to $\SS$ but not to the
natural embedding of $\RR$ into $\SS$. Validity of (3.1) in the
restricted sense ensures that this definition is consistent. \QED

Depending on the theory, the resulting net may enjoy a periodicity of
the form $B(I+N\cdot 2\pi)=B(I)$ for some $N\in\NN$, in which case it
may as well be considered as a net on the $N$-fold covering of
$S^1$. E.g., $N=1$ if $B$ is local, and $N=2$ if it is $\ZZ_2$-graded local.

To the theory on the covering, the analysis of \cite{GL,ALR} may be
applied, giving the same conclusions except those which assume that
the rotation by $2\pi$ takes an interval into itself and hence
$\Ad_{\UU(2\pi)}$ is an automorphism of $B(I)$; e.g., the
above-mentioned triviality of $\UU(4\pi)$. Thus, we have 

{\bf 3.2 Proposition \cite{GL,ALR}:} Let $I\mapsto B(I)$ be a chiral net
defined on a covering $\SS$ of the circle, satisfying the standard
assumptions: $B(I)$ are von Neumann algebras on a Hilbert space
$\HH$, $I_1\subset I_2$ implies $B(I_1)\subset B(I_2)$, there is
a unitary representation $\UU$ of $G=\widetilde{PSL}(2,\RR)$ on $\HH$
such that (3.1) holds globally on $\SS$, the rotation subgroup has a
positive generator, and there is a $\UU$-invariant vector $\Omega\in \HH$
(the vacuum) cyclic for $\bigvee_I B(I)$ and separating for $\bigcap_I
B(I)$. Then one has 

(i) {\it Reeh-Schlieder property}: $\Omega$ is cyclic and
separating for each $B(I)$. 

(ii) {\it Irreducibility and Triviality at a point}:
$\bigvee_IB(I)=\BB(\HH)$ and $\bigcap_IB(I)$ $=\bigcap_{I\ni x}B(I)=\CC\Eins$.

(iii) {\it Additivity and Continuity}: if $I$ and
$I_k$ are open intervals such that $I\subset \bigcup_k I_k$, then
$B(I)\subset \bigvee B(I_k)$, and if $\bar I$ denotes the closure of
$I$ and $\bar I\supset \bigcap_k I_k$, then $B(\bar I)\supset \bigvee B(I_k)$.

(iv) {\it Modular Covariance}: For any interval $I\subset \SS$, the modular
automorphisms $\Ad_{\Delta_I^{it}}$ of the von Neumann algebra $B(I)$
with respect to the vacuum vector $\Omega$ \cite[Chap.\ IV, Thm.\ 1.19]{T} act
geometrically by  
\begin{equation} 
\Delta_I^{it} B(J)\Delta_I^{-it} = B(\Lambda_I(-2\pi t)J) \qquad
(J\subset\SS, t\in\RR)
\end{equation}
where $\Lambda_I$ is the $I$-preserving one-parameter subgroup of $G$
which is conjugate to the scale transformations of $\RR$. 
Moreover, if $r$ is the reflection of $\SS$ induced by $x\mapsto -x$,
and $\Gamma$ the group generated by $G$ and $r$, then $\UU$ extends to
an (anti-)unitary representation of $\Gamma$ by setting
$\UU(r_I)=J_I$, where $J_I$ is the modular conjugation of
$(B(I),\Omega)$ and $r_I$ is the unique reflection in $\Gamma$
conjugate to $r$ which has the boundary points of $I$ as fixpoints.

(v) The unitaries 
\begin{equation}
z(t):=\UU(\Lambda_I(2\pi t))\Delta_I^{it}
\end{equation}
do not
depend on $I$ and form a one-parameter group in the center of the gauge
group\footnote{The gauge group consists of all unitaries $V$ on
  $\HH_B$ such that $V\Omega=\Omega$ and $VB(I)V^*=B(I)$ for all $I$.}.

(vi) {\it Bisognano-Wichmann property}: Provided $B$ is local or
$\ZZ_2$-graded local (fermionic)\footnote{This assumption will be
  substantially relaxed in the next subsection (Prop.\ 3.5).}, then 
the central cocyle $z(t)$ in (v) is trivial:
\begin{equation}
z(t)=1, \qquad\hbox{\rm i.e.,} \qquad \Delta_I^{it}=\UU(\Lambda_I(-2\pi t)). 
\end{equation}

(vii) If $z(t)=\Eins$, then one has the following equivalences:
$\CC\Omega$ are the only $\UU$-invariant vectors $\Leftrightarrow$
$B(I)$ are factors $\Leftrightarrow$ $B$ is irreducible, i.e.,
$\bigvee_I B(I)=\BB(\HH)$ $\Leftrightarrow$ $\bigcap_I B(I)=\CC\Eins$. 
In this case, if $B(I)\neq\CC\Eins$, the factors are of type $I\!I\!I_1$. 

{\em Proof:} As in \cite{GL,ALR}. (ii) is proved by various instances
of the subsequent Prop.\ 3.3: Choosing $M=\bigvee_IB(I)$ and $U$ the
subgroup of translations, gives irreducibility. Choosing
$M=\bigvee_IB(I)'$ and $U$ the translations, gives
$\bigcap_IB(I)=\CC\cdot\Eins$. Choosing $M=M_x\equiv\bigvee_{I\ni
  x}B(I)$ and $U$ the subgroup of special conformal transformations
preserving the point $x$, gives triviality at the point. (Note that
every vector which is invariant under the time translations or under
the special translations is automatically also invariant under the
full conformal group, and hence is a multiple of $\Omega$. Note also
that isotony ensures invariance of $M_x$ under the special conformal
transformations although their action does not preserve $\RR$.)\QED 

We have used

{\bf 3.3 Proposition:} Let $M$ be a von Neumann algebra on a Hilbert
space $\HH$, $v$ a cyclic vector and $U$ a one-parameter unitary group
implementing automorphisms of $M$. If $U$ has a positive generator,
and $v$ is the unique $U$-invariant vector, then $M=\BB(\HH)$. 

{\it Proof:} Let $E$ denote the projection on $\CC\cdot v$. Because
$E$ is one-dimensional and $v$ is cyclic, the algebra $E\vee M$
generated by $E$ and $M$ contains every one-dimensional projection,
hence coincides with $\BB(\HH)$. On the other hand, by positivity of
the generator \cite{Bo}, the spectrum condition implies that $U$
belongs to $M$, and consequently its spectral projection $E$ also
belongs to $M$. Hence $E\vee M$ equals $M$. \QED

For later use, we record a simple fact. 

{\bf 3.4 Proposition:} If $I\mapsto B(I)$ is a non-local net, then a
net $I\mapsto C(I)\subset B(I)$ relatively local with respect to $B$
is local. There is a unique maximal such net. The maximal net is
covariant under the covariance group of $B$, and its local algebras
$C(I)$ are globally stable under the gauge group of $B$.

{\it Proof:} Locality of $C$ is obvious. Existence and uniqueness of
the maximal net hold because any two nets $C_1$ and $C_2$ within $B$
and relatively local with respect to $B$ generate $C_1\vee C_2$ with the
same properties. The stability and covariance statements follow from
uniqueness. 

\vskip3mm
{\bf 3.2 Non-local extensions.}
\vskip2mm
In this subsection we assume that the chiral net $I\mapsto B(I)$
contains a covariant net of subfactors
$I\mapsto \pi(A(I))\subset B(I)$ which is relatively local with
respect to $B$ (in particular, $A$ is local). Then we have

{\bf 3.5 Proposition:} (i) There is a family of vacuum-preserving
conditional expectations ${\cal E}^I: B(I)\to A(I)$.

(ii) If the local subfactors $\pi(A(I))\subset B(I)$ are irreducible with
finite index, then the central cocyle (3.3) is trivial, $z(t)=1$,
i.e., the Bisognano-Wichmann property (3.4) holds. 

(iii) If $z(t)=1$, then the family of local conditional expectations
is consistent, i.e., ${\cal E}^I$ restricts
to ${\cal E}^{\hat I}$ on $B(\hat I)$ whenever $\hat I\subset I$, so
that there is a global vacuum-preserving conditional expectation
${\cal E}:B\to A$ which maps $B(I)$ onto $A(I)$. ${\cal E}$ is
implemented by the projection onto the cyclic subspace
$\overline{\pi(A)\Omega}$. 

{\it Proof:} (i) Consider the maximal net $I\mapsto C(I)$ given by 
Prop.\ 3.4. By Props.\ 3.2(v) and 3.4, $C(I)$ is globally stable 
under the modular automorphism group \cite[Chap.\ VI, Thm.\ 1.19]{T}
associated with $B(I)$ and the vacuum. By Takesaki's Theorem
\cite[Chap.\ IX, Thm.\ 4.2]{T}, there exists a vacuum-preserving
conditional expectation of $B(I)$ onto $C(I)$.  

On the other hand, because $C$ is local, we may apply \cite{ALR} or
Prop.\ 3.2(vi) to the net $C$ to conclude that  
$z(t)$ is trivial on the cyclic subspace of $C$. 
Because $\UU$ restricts to the covariance representation of $C$ which
in turn restricts to that of $A$, $A(I)$ is globally stable under the
modular automorphism group (3.4) associated with $C(I)$ and the
vacuum, so there is a vacuum-preserving conditional expectation of
$C(I)$ onto $A(I)$. Composition of the two expectations gives an
expectation ${\cal E}^I$ of $B(I)$ onto $A(I)$.  

(ii) By Prop.\ 3.2, the central cocycle $z(s)$ given by
(3.3) is a vacuum-preserving unitary one-parameter group on $\HH_B$
whose adjoint action globally preserves $B(I)$. We have to show that
$z(s)$ is trivial.  

Because there is a vacuum-preserving conditional expectation of $B(I)$
onto $A(I)$, the modular automorphisms of $B(I)$ restrict \cite{T} to the
modular automorphisms of $A(I)$ and the vacuum. Because $A$ is local,
$z(s)$ is trivial on the cyclic subspace of $A(I)$ (the vacuum
subrepresentation of $A$ in $\pi_B$). 
Hence $\Ad_{z(s)}$ is a one-parameter group of automorphisms 
of $B(I)$ acting trivially on $\pi(A(I))$. 
Thus, its fix-point subalgebra $B(I)^z$ is intermediate between
$\pi(A(I))$ and $B(I)$, and the index $[B(I):B(I)^z]$ is finite
because $[B(I):\pi(A(I))]$ is finite. But the fix-point index is the
order of the quotient group of $\RR$ by the subgroup which acts
trivially on $B(I)$. This number can be either 1 or $\infty$. 
The latter being excluded, the fix-point subalgebra must be all of
$B(I)$. Hence, the automorphic action of $\Ad_{z(s)}$ is trivial,
i.e., $z(t)$ commutes with  
$B(I)$. Since the vacuum is cyclic for $B(I)$ and $z(s)$ preserves the
vacuum, $z(s)$ itself must be trivial. This proves (ii).  

(iii) Because $z(t)=1$, the modular
automorphism group of $B\gen(I)$ coincides with the subgroup of
M\"obius transformations preserving $I$, hence it globally preserves
$A(I)$. Again by Takesaki's Theorem \cite[Chap.\ IX, Thm.\ 4.2]{T}, it
follows that ${\cal E}^I$ is implemented by the projection on the
subspace $\overline{\pi(A(I))\Omega}$. By the Reeh-Schlieder Theorem,
this projection does not depend on $I$. This implies consistency.  
\QED

In the remainder of this subsection, we shall explain the
characterization of non-local chiral extensions $I\mapsto B(I)\supset
A(I)$ in terms of Q-systems within the DHR category of the chiral net
$I\mapsto A(I)$. 

In \cite{LR}, a structural analysis of local and non-local extensions
of quantum field theories in the algebraic framework has been developed. 
The main tool was the notion of a Q-system \cite{LQ}, characterizing a
subfactor $N \subset M$ of finite index. For a brief review on
Q-systems, see App.\ A. 

A Q-system consists in a set of algebraic relations which, in the case
of quantum field theory, amount to the statement that the (non-local)
fields of the extension form a closed algebra under multiplication and
conjugation, and satisfy local commutation relations with the chiral
fields. This interpretation of a Q-system is made more transparent if
the relations are reformulated in terms of charged intertwiners (cf.\
App.\ A). 

The central result in \cite{LR} is that a Q-system $(\theta,w,x)$ 
{\em within the DHR category} of a local net $A$ determines a
relatively local net $B$ which extends $A$: $\theta$ is required to
be a DHR endomorphism of the net $O\mapsto A(O)$ localized in some 
region $O_0$, and consequently the isometries $w$ and $x$ belong to
$A(O_0)$. The Q-system therefore determines a positive-energy
representation $\pi\simeq\theta$ of the net $A$ on a Hilbert space
$\HH_B$, and the local subfactor $\pi(A(O_0))\subset B(O_0)$ on
$\HH_B$. The latter can then be ``transported'' to a covariant net of
subfactors $O\mapsto[\pi(A(O))\subset B(O)]$ equipped with a
consistent family of conditional expectations preserving the
vacuum. (Imposing the additional eigenvalue condition
$\eps(\theta,\theta) x=x$ would ensure $O\mapsto B(O)$ to be a 
{\em local} net.)  

In the case of (non-local) chiral
extensions $A(I)\subset B(I)$ at hand, $A$ being completely rational
implies that only finitely many (equivalence classes of) endomorphisms
$\theta$ can appear in an irreducible Q-system: the argument is as in
\cite[Prop.\ 2.3]{KL}, using the fact that the multiplicity $n_s$ of
each irreducible subsector $[\rho_s]$ of $\theta$ is bounded by the
square of its dimension \cite[p.\ 39]{ILP}.  In particular, the index
of the local subfactors $A(I)\subset B(I)$ is finite (and so the stronger
bound $n_s\leq d(\rho_s)$ \cite[Cor.\ 4.6]{LR} applies). Moreover, it
was shown in \cite[Thm.\ 2.4]{IK} that each $\theta$ can arise only in
finitely many inequivalent Q-systems. This means that the classification
problem of Q-systems in the DHR category of a (completely) rational
CFT is a finite problem with finitely many solutions, and thus, fixing
$A$, there exist only finitely many non-local chiral extensions $B$.  

Examples for Q-systems within the DHR category of a local net were given
for local and non-local chiral extensions of chiral nets, and for
local two-dimensional extensions of subnets $A_L\otimes A_R$
consisting of two (left and right) chiral nets \cite{LR}. The main
result in \cite{CTPS} is that there is a systematic way (the 
{\em $\alpha$-induction construction}, using results of \cite{BEK1}) to
associate a local Q-system, and hence a local two-dimensional
extension $B^\alpha_2$ of $A_2 = A\otimes A$, with any given chiral
extension $B$ of $A$.  
\vskip3mm
{\bf 3.3 The split property.}
\vskip2mm
Let us now turn to the split property, which is related to phase space
properties (existence of $\Tr\exp-\beta L_0$) in QFT \cite{BAL,ALR}. 
A commuting pair of von Neumann algebras $(M_1,M_2)$ is split if there
is a natural isomorphism from $M_1\vee 
M_2$ to $M_1\otimes M_2$, where $M_1\vee M_2$ denotes the von Neumann
algebra generated by $M_1$ and $M_2$. A chiral net $B$ is split if the
pair $(B(K),B(L)')$ is split whenever the open interval $L$ contains
the closure of the interval $K$. We want to prove ``upward
hereditarity'' of the split property. 

{\bf 3.6 Proposition:} Let $B$ be a M\"{o}bius covariant net on $S^1$
and $A$ a finite index subnet such that $B$ is relatively local with  
respect to $A$. If $A$ is split, then also $B$ is split.

Note that $B$ is possibly non-local, but relative locality implies
that $A$ is local. In the case of a {\em local} net
$B$, the result was proven in \cite{L}. 

To prepare the proof of Prop.\ 3.6, we need Prop.\ 3.7 and Lemma 3.8:
\newpage
{\bf 3.7 Proposition:}
Let $M_1$, $M_2$ be commuting factors and $N_k\subset M_k$ finite index 
subfactors, $k=1,2$. If the pair $(N_1,N_2)$ is split, then also the pair 
$(M_1,M_2)$ is split.

Let $\gamma_k:M_k\to N_k$ be canonical endomorphisms and 
$(\gamma_k,T_k,S_k)$ the associated Q-systems. Then $M_k = N_k T_k$,
and every $m^{(k)}\in M_k$ can be written as $m^{(k)}= n^{(k)}T_k$ where  
$n^{(k)}\in N_k$ is given by $n^{(k)}=\lambda_k\cdot{\cal E}_k(m^{(k)}T^*_k)$, 
with ${\cal E}_k$ the associated expectation from $M_k$ to $N_k$ and
$\lambda_k$ is the index $[M_k:N_k]$. Thus 
$||n^{(k)}||\leq\lambda_k ||m^{(k)}||$.  

{\bf 3.8 Lemma:} With the above notations we have $NT_1 T_2 = M$ where 
$M=M_1\vee M_2$, $N=N_1\vee N_2$.

Moreover there is a constant $C>0$ such that if $m\in M$ then 
$m=nT_1 T_2$, with $n\in N$ and $||n||\leq C||m||$.

{\it Proof of Lemma 3.8:} We first show that the second part of the
statement with $m\in M_1\cdot M_2$. Here $M_1\cdot M_2$ is the product 
of $M_1$ and $M_2$ which is naturally isomorphic to the algebraic tensor 
product $M_1\odot M_2$ by the Murray-von Neumann factorization lemma.

Let $m=\sum m^{(1)}_i m^{(2)}_i$ with $m^{(1)}_i\in M_1, m^{(2)}_i\in 
M_2$ and write $m^{(1)}_i= n^{(1)}_i T_1$, $m^{(2)}_i= n^{(2)}_i T_2$, with 
$n^{(k)}_i \in N_k$. 

The subfactor $N_1\otimes N_2$ of $M_1\otimes M_2$ has finite index 
and the associated Q-system is the tensor product Q-system, hence 
there is a constant $C>0$ such that $||\sum n^{(1)}_i\otimes n^{(2)}_i||\leq 
C||\sum m^{(1)}_i\otimes m^{(2)}_i||$ where the norms here are the 
spatial tensor product norms. Hence we have
\begin{equation}
||\sum n^{(1)}_i n^{(2)}_i||= ||\sum n^{(1)}_i\otimes n^{(2)}_i||
\leq C||\sum m^{(1)}_i\otimes m^{(2)}_i||\leq 
C||\sum m^{(1)}_i m^{(2)}_i||,
\end{equation}
where the first equality holds because of the split property for 
$(N_1,N_2)$ and the last inequality due to the minimality of the spatial 
tensor product norm.

Now we prove the general statement. Let $m\in M$ with $||m||\leq 1$
and choose by Kaplanski density theorem a net of elements $m_j\in 
M_1\cdot M_2$,  With $||m_j||\leq 1$ and $m_j\to m$ weakly. 

We can write $m_j = n_j T_1 T_2$ where $n_j\in N$ and $||n_j||\leq C$.
With $n$ a weak limit point of $n_j$, we then have $m=n T_1 T_2$ 
and $||n||\leq C$. \QED

{\it Proof of Prop.\ 3.7:} 
Let $\Phi: M_1\otimes M_2 \to M$ be 
the linear map 
\begin{equation}
m\mapsto \Phi(m)\equiv \Phi_0(n)T_1 T_2
\end{equation}
where $n\in N_1\otimes N_2$ is the unique element such that $m= n\cdot
T_1\otimes T_2$ and $\Phi_0=\Phi\rest_{N_1\otimes N_2}$ is the
natural isomorphism of $N_1\otimes N_2$ with $N_1\vee N_2$. By the
Lemma, $\Phi$ is surjective.

We show that $\Phi$ is multiplicative and respects the $*$
operation. First note that if $n\in N$ then
\begin{equation}
T_1 T_2 n =\theta(n)T_1 T_2
\end{equation}
where $\theta$ is the endomorphism of $N$ which is transformed to 
$\gamma_1\rest_{N_1}\otimes\gamma_2\rest_{N_2}$ under 
$\Phi_0$ (check this with $n\in N_1\cdot N_2$, then it holds for 
all $n\in N$ by continuity).

Let $m'\in M_1\otimes M_2$, $m'= n'\cdot T_1\otimes 
T_2$ with $n'\in N_1\otimes N_2$. Then 
\begin{equation}
\begin{array}{r}
mm' = n\cdot T_1\otimes T_2 \cdot n'\cdot T_1\otimes T_2 
=  n\gamma_1\otimes\gamma_2(n')\cdot T^2_1\otimes T^2_2 =\\=
n\gamma_1\otimes\gamma_2(n')\cdot \lambda_1{\cal E}_1(T^2_1 T^*_1)T_1
\otimes \lambda_2{\cal E}_2(T^2_2 T^*_2)T_2\ ,
\end{array}
\end{equation}
Thus, suppressing the symbol $\Phi_0$ for simplicity,
\begin{equation}
\Phi(mm')=n\theta(n')\cdot \lambda_1{\cal E}_1(T^2_1 T^*_1)
\lambda_2{\cal E}_2(T^2_2 T^*_2) \cdot T_1 T_2=n\theta(n')T^2_1 T_2^2\ .
\end{equation}
On the other hand
\begin{equation}
\Phi(m)\Phi(m')= nT_1 T_2 n'T_1 T_2=n\theta(n')T^2_1 T_2^2
\end{equation}
as desired. As for the $*$ operation, the argument is completely
analogous, using the formula $T_i{}^*=\lambda_i{\cal E}_i(T_i{}^*{}^2)T_i$.

Thus, $\Phi$ is a $^*$-homomorphism of von Neumann 
algebras, hence $\sigma$-weakly continuous.
Since $M_1\otimes M_2$ is a factor, $\Phi$ is injective and 
$(M_1, M_2)$ is a split pair.
\QED

{\it Proof of Prop.\ 3.6.} Let $I\subset\subset \tilde I$ be intervals
and apply Prop.\ 3.7 with $N_1 = A(I)$, $N_2 = A(\tilde I')$, 
$M_1 = B(I)$, $M_2=B(\tilde I)'$. We just note that $[M_2 : N_2] <
\infty$ because the inclusion $A(\tilde I')\subset B(\tilde I)'$ is
anti-isomorphic to  
\begin{equation}
J_{\tilde I}A(\tilde I')J_{\tilde I}=A(\tilde I)\subset 
J_{\tilde I}B(\tilde I)'J_{\tilde I}=B(\tilde I)
\end{equation}
where $J_{\tilde I}$ is the modular conjugation of $B(\tilde I)$ with
respect to the vacuum and we are using the geometric action of $J_{\tilde I}$. 
\QED

\section{Charged intertwiners in boundary CFT}
\setcounter{equation}{0} 

The main result of this section is a generalization of 
\cite[Thm.\ 3.1 and Remark 3.2]{K} (where $B$ was assumed to be local): 

{\bf 4.1 Theorem:} For a given completely rational chiral net 
$I\mapsto A(I)$, and a given irreducible (possibly non-local) chiral
extension $I\mapsto B(I)$, let $O\mapsto B\ind_+(O)$, $O=I\times J
\subset M_+$, be the induced Haag dual boundary CFT net (Def.\ 2.8),
and let $O\mapsto B^\alpha_2(O)$, $O=I\times J \subset M$, be the
two-dimensional local net on Minkowski space extending $A\otimes A$,
obtained from $B$ by the $\alpha$-induction construction. Then the
local subfactors 
\begin{equation}
A(I)\vee A(J)\subset B\ind_+(O)\qquad\hbox{and}\qquad A(I)\otimes
A(J)\subset B^\alpha_2(O)
\end{equation} 
are isomorphic.  

Because $B_+(O)$ is intermediate between $A_+(O)$
and the dual net $B\dual_+(O)$, we conclude 

{\bf 4.2 Corollary:} For a given boundary  CFT $O\mapsto B_+(O)$,
let $O\mapsto B^\alpha_2(O)$ be the local two-dimensional extension  
of $A\otimes A$ obtained by applying the $\alpha$-induction
construction to the boundary net $B$ of $B_+$. Under the isomorphism
established in Thm.\ 4.1, the net $O\mapsto B_+(O)$ corresponds to an
intermediate net 
\begin{equation}
A(I)\otimes A(J)\subset B_2(O)\subset B^\alpha_2(O),
\end{equation} 
with $B_2(O)=B^\alpha_2(O)$ if and only if $B_+$ satisfies Haag duality.  

In the course of the proof of the theorem, we shall ``compute'' the
relative commutant $B\ind_+(O)=B(L)\cap B(K)'$ by determining local
operators $\psi_i\in B\ind_+(O)$ ({\em charged intertwiners}, see
below) which along with $A_+(O)$ generate $B\ind_+(O)$. These charged
intertwiners, as $O$ varies, are the von Neumann analog of the
non-chiral local Wightman fields of the boundary CFT, generalizing
(1.11), (1.12). In Sect.\ 5, we shall further analyse their
bi-localized charge structure.  

In a general subfactor setup (see App.\ A for more details), the
charged intertwiners for a subfactor $N\subset M$ 
are nontrivial elements $\psi_i$ of $M$ satisfying 
\begin{equation}
\psi_i n=\varrho_i(n)\psi_i \qquad (n\in N)
\end{equation}
(where $\varrho_i$ are irreducible endomorphisms of $N$), such that 
every element of $M$ has a unique expansion 
\begin{equation}
m = \sum_i n_i\psi_i,\qquad (n_i\in N).
\end{equation}

The {\em algebra of the charged intertwiners} in $M$ 
\begin{equation}
\psi_i\psi_j= \sum_k \Gamma_{ij}^k\, \psi_k, \qquad 
\psi_i^* = \sum_j \Gamma_{ji}^{0*}\,\psi_j.
\end{equation}
with intertwiners $\Gamma_{ij}^k:\varrho_k\to\varrho_i\varrho_j$ in $N$, 
together with some normalization conditions, is an invariant of the
subfactor $N\subset M$, determined by the Q-system. 

Generalizing an argument used in \cite{KLM}, we show in Lemma A.2 that
the Q-systems $(\gamma,v,w)$ in $M$ and $(\theta,w,x)$ in $N$ can be
recovered from the system of charged intertwiners $\psi_i\in M$; in 
particular, two such systems $\psi_i\in M$, $\tilde\psi_i\in \tilde M$
satisfying the same algebra with the same $\varrho_i\in\End(N)$ and
$\Gamma^k_{ij}\in N$ induce an isomorphism of subfactors $N\subset M$,
$N\subset \tilde M$ by $\psi_i\leftrightarrow \tilde\psi_i$. 

Applying this argument to $N=A(I)\vee A(J)\simeq A(I)\otimes A(J)$ and
$M=B\ind_+(O)$, $\tilde M=B^\alpha_2(O)$, the statement of the theorem
thus follows from the equivalence of the algebras of charged
intertwiners which generate the respective inclusions.  

{\it First part of the proof of the Theorem:} We proceed in close
analogy with the proof of \cite[Prop.\ 45]{KLM}. $A_2(O)=A(I)\otimes
A(J)$ and $A_+(O)= A(I)\vee A(J)$ are naturally isomorphic by the
split property of the chiral net $A$. Under this isomorphism, the
Q-system $(\Theta_2,W_2,X_2)$ for $A_2(O)\subset B^\alpha_2(O)$ (given
in \cite{CTPS}, see below) turns into a Q-system $(\Theta,W,X)$ in
$A_+(O)$ with 
\begin{equation}
X = d(\Theta)^{-\frac12}\sum_{ijk}
W^{i}\Theta(W^j)\,\Gamma_{ij}^k\, W^{k*}.
\end{equation}
By the preceding discussion and Lemma A.2, it is sufficient to show
that $\Theta$ coincides with the dual canonical endomorphism $\Theta_+$ 
for $A_+(O)\subset B\ind_+(O)$, and to find charged intertwiners
$\psi_i\in B\ind_+(O)$ satisfying the algebra (A.4--6+9) with
$\Gamma_{ij}^k$ given by (4.10). 

Without loss of generality, we choose $I=(y,z)\subset \RR_+$ 
and $J=-I \subset \RR_-$ (this situation can always be attained by a 
conformal transformation), thus $K=(-y,y)$, $L=(-z,z)$ are symmetric, and put 
$O=I\times J$. Then $A(I)=j(A(J))$ where $j=\Ad_J$ is the modular
conjugation \cite[Chap.\ IV, Thm.\ 1.19]{T} for $A(\RR_+)$ with respect to the
vacuum (= PCT transformation \cite{GL}), and $A_+(O)=A(I)\vee j(A(I))$.  

We choose a system $\Delta$ of inequivalent irreducible DHR endomorphisms 
$\rho_s$ localized in $I$, thus $\bar\rho_s=j\circ\rho_s\circ j$ are
conjugates of $\rho_s$ localized in $J$. We have

{\bf 4.3 Lemma:} Every irreducible subsector of $\Theta_+$ is equivalent to 
some $\sigma\bar\tau$ with $\sigma,\tau\in\Delta$.

The proof of this Lemma is exactly as the proof of \cite[Lemma 31]{KLM}.

Next, we show an analog of \cite[Theorem 9]{KLM}, which implies that
$\Theta\simeq\Theta_2$ indeed coincides with the dual canonical
endomorphism $\Theta_+$ for $A_+(O)\subset B_+\ind(O)$: 

{\bf 4.4 Proposition:} The multiplicities of $[\sigma\bar\tau]$ in 
the dual canonical endomorphism $\Theta_+$ for $A_+(O)\subset B\ind_+(O)$
equal $Z_{[\sigma][\tau]}=\dim \Hom(\alpha^+_\tau,\alpha^-_\sigma)$,
where $\alpha^\pm_\rho$ are the $\alpha$-induced extensions of $\rho\in
\Delta$ to the chiral net $B$ (see App.\ B).

{\it Proof:} By Def.\ B.1, $\alpha^\pm_\rho$ are endomorphisms of
$B(L)$, and by Prop.\ B.3, the global intertwiners coincide with the 
local intertwiners, i.e., $t\alpha^+_\tau(b) = \alpha^-_\sigma(b)t$
holds for all $b\in B$ iff it holds for all $b\in B(L)$, and in this
case $t$ belongs to $B(L)$.

Now, for $\sigma,\tau\in\Delta$, consider the space $X_{\sigma\tau}$ of 
intertwiners $\psi\in B\ind_+(O)$ satisfying
\begin{equation}
\psi a = \sigma\bar\tau(a)\psi \qquad (a\in A_+(O)).
\end{equation}
Then for $\psi\in X_{\sigma\tau}$, the same equation (4.7) also holds
with $a\in A(K)$ and $a\in A(L')$ because then $\sigma\bar\tau(a)=a$
and because $B\ind_+(O)$ commutes with $A(K)$ and $A(L')$. By strong
additivity of $A$, (4.7) holds in fact for all $a\in A$.

Consider on the other hand the space $\Hom(\alpha^+_\tau,\alpha^-_\sigma)$ 
of intertwiners $t\in B$ satisfying
\begin{equation}
t\alpha^+_\tau(b) = \alpha^-_\sigma(b)t \qquad (b\in B)
\end{equation}
whose dimension is $Z_{[\sigma][\tau]}$. We claim that the map
\begin{equation}
\varphi: t\mapsto\psi:= t R_\tau
\qquad
\varphi\inv: \psi\mapsto t:= \sigma(\bar R_\tau^*) \psi
\end{equation}
is an isomorphism between $\Hom(\alpha^+_\tau,\alpha^-_\sigma)$ and
$X_{\sigma\tau}$, which proves the proposition. Here, 
$R_\tau:\id\to\tau\bar\tau$ are the standard intertwiners in $A(L)$ as
in \cite{GL,KLM}, normalized such that $R_\tau^*R_\tau=d(\tau)$, and 
$\bar R_\tau= \kappa_\tau\cdot R_\tau$ such that by \cite[Proof of
Lemma 3.5]{GL} one has $\tau(\bar R^*)R = \bar R^*\bar\tau(R)=1.$
The latter normalization condition ensures that the maps (4.9) are
mutually inverse. It remains to show that the image of 
$\Hom(\alpha^+_\tau,\alpha^-_\sigma)$ belongs to $X_{\sigma\tau}$, and
vice versa. 

Let $t\in\Hom(\alpha^+_\tau,\alpha^-_\sigma)$. Then $t\in B(I)$ by
Prop.\ B.3, and $\psi:=\varphi(t)=tR_\tau$ by definition belongs to
$B(L)$ and satisfies (4.7) for all $a\in A$. The non-trivial part is
to show that it also belongs to $B(K)'$. Because $\sigma\bar\tau$ acts
trivially on $A(K)$, $\psi$ commutes with $A(K)$ by (4.7). Because $A(K)$
and $v$ generate $B(K)$, where $(\gamma,v,w)$ is the Q-system for
$A(K)\subset B(K)$, it suffices to show that $\psi$ commutes with 
$v\in B(K)$. We compute (with Prop.\ B.4(i)) 
\begin{equation}
\psi v = tR_\tau v = t\alpha^+_\tau\alpha^+_{\bar\tau}(v)R_\tau v =
t\alpha^+_\tau(v)R_\tau = \alpha^-_\sigma(v)tR_\tau = vtR_\tau = v\psi,
\end{equation}
because $\alpha^+_\tau$ and $\alpha^-_\sigma$ act trivially on $v\in B(K)$. 
Thus, $\psi\in B(K)'$ and  
$\varphi(\Hom(\alpha^+_\tau,\alpha^-_\sigma))\subset X_{\sigma\tau}$.

Conversely, let $\psi\in X_{\sigma\tau}$. By definition,
$t:=\varphi\inv(\psi)$ belongs to $B(L)$ and satisfies
\begin{equation} 
t\tau(a) = \sigma(a)t \qquad (a\in A(L)).
\end{equation}
Thanks to Prop.\ B.3, it remains to show that $t$ has the required
intertwining property
\begin{equation}
t\alpha^+_\tau(v) = \alpha^-_\sigma(v)t.
\end{equation}
Inserting the above definitions for $t=\varphi\inv(\psi)$ and for
$\alpha^\pm_\rho(v)$, we have
\begin{equation}
t\alpha^+_\tau(v) = \sigma(\bar R_\tau^*)\psi\eps(\theta,\tau)v = 
\sigma(\bar R_\tau^*)\sigma\bar\tau(\eps(\theta,\tau))\psi v =
\sigma(\eps(\theta,\bar\tau)^*)\sigma\theta(\bar R_\tau^*)\psi v, 
\end{equation}
where $\theta=\gamma\rest_A$ is localized in $K$, and 
\begin{equation}
\alpha^-_\sigma(v)t = \eps(\sigma,\theta)^*v\sigma(\bar R_\tau^*)\psi 
= \eps(\sigma,\theta)^*\theta\sigma(\bar R_\tau^*)v\psi 
= \sigma\theta(\bar R_\tau^*)\eps(\sigma,\theta)^*v\psi.
\end{equation}
In (4.13), the statistics operator $\eps(\theta,\bar\tau)$ is trivial
because of the ordering $J<K$ of the localizations of the
endomorphisms, and in (4.14) $\eps(\sigma,\theta)$ is trivial because
$K<I$. Because $\psi\in X_{\sigma\tau}$ belongs to $B(K)'$ and 
$v\in B(K)$, $\psi$ commutes with $v$, hence (4.13) and (4.14) are
equal. Thus (4.8) holds, and 
$\varphi\inv(X_{\sigma\tau})\subset \Hom(\alpha^+_\tau,\alpha^-_\sigma)$.
This completes the proof of the proposition. \QED

{\it Proof of the Theorem (continued):} The dimensions
$Z_{[\sigma][\tau]}$ in the Prop.\ 4.4 being the 
multiplicities of $[\sigma\bar\tau]$ in $\Theta$, we conclude
$\Theta_+=\Theta\simeq \bigoplus Z_{[\sigma][\tau]}\sigma\bar\tau$. 
For each pair $\sigma$, $\tau\in\Delta$ such that
$\sigma\bar\tau\prec\Theta_+$, we fix a basis of charged intertwiners 
\begin{equation}
\psi_i:=\varphi(t_i) = t_iR_{\tau_i}
\end{equation}
where $t_i$ are bases of the spaces $\Hom(\alpha^+_{\tau},\alpha^-_{\sigma})$  
orthonormal with respect to their inner products $\langle t,t'\rangle
= (d(\sigma)d(\tau))\inv\cdot R^*_{\tau} t^*t'R_{\tau}$. As $\sigma$
and $\tau$ vary over $\Delta$, we thus obtain a maximal system of
charged intertwiners $\psi_i$ in $B\ind_+(O)$, $i=1\ldots\sum
Z_{[\sigma][\tau]}$, normalized as
\begin{equation}
\psi_i^*\psi_j=d(\sigma)d(\tau)\cdot \delta_{ij}.
\end{equation}
We claim that these form an algebra of charged
intertwiners with endomorphisms\footnote{The index $i$ thus labels the
  irreducible components of $\Theta_+\in\End(A_+(O))$, which may be
  pairwise equivalent whenever $Z_{[\sigma_i][\tau_i]}>1$.} 
$\varrho_i=\sigma_i\bar\tau_i\prec \Theta_+$ and coefficients
$\Gamma^k_{ij}$ as in (4.6), i.e., those of the $\alpha$-induction
construction given more explicitly in (4.17) below. Let us recall how
the latter  were determined.  

The $\alpha$-induction construction \cite{CTPS} proceeds by the
specification of a Q-system in $A(I)\otimes A(I)\opp$, which under the
isomorphism between $A(I)\opp$ and $A(J)$ given by $a\opp\mapsto j(a^*)$ 
turns into the Q-system $(\Theta_2,W_2,X_2)$ in $A(I)\otimes A(J)$, 
determining the extension $A(I)\otimes A(J)\subset B^\alpha_2(O)$ up
to isomorphism. Applying in turn the natural isomorphism 
$A(I)\otimes A(J)\simeq A(I)\vee A(J)$, we read off \cite[Sect.\ 3]{CTPS} 
\footnote{adapting the notation of \cite{CTPS} to the present conventions.}
\begin{equation}
\Gamma_{ij}^k = d(\Theta)^{\frac12}\sum_{ef} \zeta_{ij,ef}^k \cdot
T_e\;j(T_f) 
\in\Hom(\varrho_{k},\varrho_{i}\varrho_{j})
\end{equation}
where $\varrho_i=\sigma_i\bar\tau_i\prec\Theta$, $T_e$ form orthonormal
bases of $\Hom(\sigma_k,\sigma_i\sigma_j)\subset A(I)$, $T_f$ form
orthonormal bases of $\Hom(\tau_k,\tau_i\tau_j)\subset A(I)$  
and consequently $j(T_f)$ form orthonormal bases of
$\Hom(\bar\tau_k,\bar\tau_i\bar\tau_j)\subset A(J)$, and the numerical
coefficients $d(\Theta)^{\frac12}\cdot\zeta_{ij,ef}^k$ 
are the expansion coefficients of
\begin{equation}
t_i\;\alpha^+_{\tau_i}(t_j)\in 
\Hom(\alpha^+_{\tau_i\tau_j},\alpha^-_{\sigma_i\sigma_j})
\end{equation}
into the basis $T_e\,t_k\,T_f^*$
of $\Hom(\alpha^+_{\tau_i\tau_j},\alpha^-_{\sigma_i\sigma_j})$:
\begin{equation}
t_i\alpha^+_{\tau_i}(t_j)=
d(\Theta)^{\frac12}\sum_{k,ef}\zeta_{ij,ef}^k\cdot T_e\,t_k\,T_f^*.
\end{equation}

We now compute
\begin{equation}
\psi_i\psi_j= t_iR_{\tau_i}\cdot t_jR_{\tau_j}= 
t_i\alpha^+_{\tau_j}\alpha^+_{\bar\tau_j}(t_j)\cdot
R_{\tau_i}R_{\tau_j}= t_i\alpha^+_{\tau_j}(t_j)\cdot
R_{\tau_i}R_{\tau_j},
\end{equation}
because $\alpha^+_{\bar\tau_j}$ acts trivially on $B(I)$, and insert
(4.19) as well as \cite[Eq.\ (15)]{KLM}
\begin{equation}
R_{\tau_i}R_{\tau_j}=\sum_{g}T_g\,j(T_g)\cdot R_{\tau_k},
\end{equation}
$T_g\in\Hom(\tau_k,\tau_i\tau_j)$.
This yields
\begin{equation}
\psi_i\psi_j= d(\Theta)^{\frac12}\sum_{k,ef}\zeta_{ij,ef}^k\cdot
T_e\,t_k\,j(T_f) \cdot R_{\tau_k} = \Gamma^k_{ij}\cdot \psi_k
\end{equation}
because $t_k\in B(I)$ and $j(T_f)\in A(J)$ commute.

It remains to prove the second of the two defining relations (A.9) 
with $\Gamma_{ji}^0$ determined by (4.17). We observe that by
definition of $\Gamma_{ji}^0$ only $\tau_j$ conjugate to $\tau_i$ and
$\sigma_j$ conjugate to $\sigma_i$ contribute, and the sums over $e$
and $f$ involve only one term $T_e\in\Hom(\id,\sigma_j\sigma_i)$ and
$T_f\in\Hom(\id,\tau_j\tau_i)$. If we prove (the first of) the identities
\begin{equation}
\begin{array}{l}
d(\Theta)^{\frac12}\sum_{j}\zeta_{ji,ef}^0\cdot t_j^*T_e =
d(\tau_j)\cdot \alpha_{\tau_j}^+(t_i)T_f, \\ 
d(\Theta)^{\frac12}\sum_{j}\zeta_{ji,ef}^0\cdot T_f^*t_j^* =
d(\sigma_j)\cdot T_e^*\alpha_{\sigma_j}^-(t_i).
\end{array}
\end{equation}
then the claim reduces to the corresponding result obtained in
\cite[eqs.\ (11) and (12)]{KLM}: namely we get (using local
commutativity of $t_j\in B(J)$ with $j(T_f)\in A(I)$ as well as  
$j(T_f^*)R_{\tau_j}=\tau_j(j(T_f^*))R_{\tau_j}\in\Hom(\bar\tau_i,\tau_j)$
and the trivial action of $\alpha_{\bar\tau_i}^+$ on $t_i$ due to
Prop.\ B.2)
\begin{equation}
\begin{array}{r}
\sum_j\Gamma_{ji}^{0*}\psi_j = 
d(\Theta)^{\frac12}\sum_{j}\overline{\zeta_{ji,ef}^0}\cdot T_e^*t_j
j(T_f^*)R_{\tau_j} \stackrel{\rm (4.23)}= \qquad\qquad\qquad\\ = d(\tau_j)\cdot T_f^*
\alpha_{\tau_j}^+(t_i^*)j(T_f^*)R_{\tau_j} \stackrel{\rm Prop.\ B.4}=
d(\tau_j)\cdot T_f^* j(T_f^*)R_{\tau_j}\alpha_{\bar\tau_i}^+(t_i^*)= \\
\stackrel{\hbox{\small\cite{KLM}}}= R_{\tau_i}^*\alpha_{\bar\tau_i}^+(t_i^*)
\stackrel{\rm Prop.\ B.2}= R_{\tau_i}^*t_i^* = \psi_i^*.
\end{array}
\end{equation}
To prove (4.23), we choose $T_g\in\Hom(\id,\tau_i\tau_j)$ such that 
$\tau_i(T_f^*)T_g=\Eins$, and consequently also
$T_f^*\tau_j(T_g)=\Eins$. Let $\tilde t:=
[T_e^*\alpha_{\sigma_j}^-(t_iT_g)]^* \in\Hom(\tau_i,\sigma_i)$. Then
\begin{equation}
d(\Theta)^{\frac12}\zeta_{ji,ef}^0 =
T_e^*\alpha_{\sigma_j}^-(t_i)t_jT_f
= T_f^*\tilde t^*t_jT_f = d(\tau_i)\inv R_{\tau_j}^*\tilde
t^*t_jR_{\tau_j} = d(\sigma_j)\langle\tilde t,t_j\rangle,
\end{equation}
hence $\sum_jd(\Theta)^{\frac12}\zeta_{ji,ef}^0 \cdot t_j^* =
d(\sigma_j)\tilde t^*$ because $t_j$ form an orthonormal basis with
respect to $\langle\cdot,\cdot\rangle$. Inserting the definitions, we get
\begin{equation}
\begin{array}{r}
d(\Theta)^{\frac12}\sum_j\zeta_{ji,ef}^0 \cdot t_j^*T_e=d(\sigma_j)\cdot
\tilde t^*T_e = d(\sigma_j)\cdot T_e^*\alpha_{\sigma_j}^-(t_iT_g)T_e
=\qquad\qquad 
\\ = d(\tau_j)\cdot T_f^*\alpha_{\tau_j}^+(T_gt_i)T_f
= d(\tau_j)\cdot T_f^*\tau_j(T_g)\alpha_{\tau_j}^+(t_i)T_f
= d(\tau_j)\cdot \alpha_{\tau_j}^+(t_i)T_f,
\end{array}
\end{equation}
using the fact that $T_e$ and $T_f$ implement standard left-inverses
for the $\alpha$-induced sectors, and the trace property for standard
left-inverses. Similarly, 
\begin{equation}
\begin{array}{r}
d(\Theta)^{\frac12}\sum_j\zeta_{ji,ef}^0 \cdot T_f^*t_j^*=d(\sigma_j)\cdot
T_f^*\tilde t^* = d(\sigma_j)\cdot
T_f^*T_e^*\alpha_{\sigma_j}^-(t_iT_g) =
\\ = d(\sigma_j)\cdot T_e^*\alpha_{\sigma_j}^-(t_i\tau_i(T_f^*)T_g)
= d(\sigma_j)\cdot T_e^*\alpha_{\sigma_j}^-(t_i),
\end{array}
\end{equation}
proving (4.23).

This completes the proof of Theorem 4.1. \QED

{\it Proof of Corollary 4.2:}  Obvious. 

{\it Remark:} In 2D CFT, there is a pair of maximal left and
right chiral algebras such that $A_L(I)\otimes A_R(J)\subset 
A\max_L(I)\otimes A\max_R(J)\subset B_2(O)$. Under standard
assumptions \cite{CO}, these are given by $A\max_L(I)=B_2(I\times J)
\cap (\Eins\otimes A_R(J))'$ (independent of $J$) and similar for
$A\max_R$. In the present situation, with $A_L=A_R=A$ and
$B_2=B^\alpha_2$, the isomorphism of Thm.\ 4.1 identifies
$A\max_L(I)$ with $B(L)\cap B(L\setminus\bar I)$.  
Namely, $B\ind_+(O)\cap A(J)'=B(L)\cap (B(K)\vee A(J))'$ and $B(K)\vee
A(J)= \{v^K\}\vee A(K)\vee A(J)=\{v^K\}\vee A(L\setminus \bar I) = 
B(L\setminus \bar I)$ by strong additivity of $A$. In particular, 
the intersection $B(L)\cap B(L\setminus\bar I)$ does not depend on the
upper boundary of the interval $L$ and may be replaced by 
$\hat A_L(I):=B((a,\infty))\cap B((b,\infty))'$ if $I=(a,b)$. 

The chiral nets $I\mapsto \hat A_L(I)$ and $I\mapsto \hat A_R(I):=
B((-\infty,b))\cap B((-\infty,a))'$ thus define two local and
mutually local chiral nets, both extending $I\mapsto A(I)$ within
$B(I)$, such that $A(I)\vee A(J)\subset \hat A_L(I)\vee \hat A_R(J)\subset
B\ind_+(O)$ for $J<I$. In the setting of \cite{BE2}, they correspond
to the intermediate subfactors $N\subset M_\pm\subset M$.

\section{Bi-localized charge structure in BCFT}
\setcounter{equation}{0} 
Our aim in this section is to establish in the algebraic framework
formulae of the type (1.11), (1.12), exhibiting a separation of the left and
right charges of local fields in BCFT
({\it bi-localized charge structure}). This will explain Cardy's
observation \cite{C} concerning the relation between $n$-point local
correlation functions and $2n$-point conformal blocks in a
model-independent setting. Furthermore, it enables us to compute the
specific linear coefficients which guarantee locality, in terms of the
DHR structure of the underlying net $A$ of chiral observables.
 \vskip3mm
{\bf 5.1. Preliminaries.}
\vskip2mm
Let us recall and adapt for our present purposes several results from
the literature.  

In \cite{DHR}, under the name of {\em field bundle} a ``crossed
product action of the DHR category on the observables'' has been
constructed as a first substitute for an algebra of charged fields has
been constructed. The fibres of this bundle were labelled by all the DHR
endomorphisms. The huge redundancy has been eliminated with the
``reduced field bundle'' in \cite{FRS1,FRS2} where only one fibre was
retained for each irreducible superselection sector. 

This amounts to a choice, for each irreducible sector $[s]$, of a
representative DHR endomorphism $\rho_s$ along with the representation
of the observables on the Hilbert space $\HH_s$. As a {\em space},
$H_s$ coincides with the vacuum Hilbert space $\HH_0$ of the net $A$,
but as a {\em representation} it differs in that $A$ is represented on
$\HH_s$ under the action of the endomorphism $\rho_s$, i.e., $\pi_s=\rho_s$. 
We call $\hat\HH$ the direct sum of the $\HH_s$ (which is finite
because $A$ is rational), and $\hat\pi$ the corresponding
representation. 

Let $\sigma$ be a DHR endomorphism of $A$ and $T_e$ an orthonormal
basis of intertwiners $T_e:\rho_s\to\rho_t\sigma$, $e=1\dots\dim
\Hom(\rho_s,\rho_t\sigma)$. Then $T_e$, as an operator from $\HH_s$ to
$\HH_t$, satisfies the intertwining relation
\begin{equation}
T_e\pi_s(a)=\pi_t(\sigma(a))T_e.
\end{equation}
It is crucial that $T_e$, although an element of $A$ as an operator,
must not be considered as an {\em observable} since it acts
on $\HH_s$ in the representation $\pi_0=\id$, and not in the
representation $\pi_s$ pertaining to $\HH_s$. We emphasize this fact 
by our notation, and denote by $\psi^\sigma_e$ the operator on
$\hat\HH$ which coincides with $T_e$ on the subspace $\HH_s$ (with
values in $\HH_t$) and is extended by zero on its orthogonal
complement in $\hat\HH$.\footnote
  {The same operator was denoted $F_e(1)^*$ in \cite{FRS2}.} Thus 
\begin{equation}
\psi^\sigma_e\hat\pi(a)=\hat\pi(\sigma(a))\psi^\sigma_e.
\end{equation}
If $\sigma$ is localized in an interval $I$, then $\sigma(a)=a$ for
$a\in A(I')$, hence $\psi^\sigma_e$ commutes with $\hat\pi(A(I'))$. We
therefore arrive at the ``reduced field net'' \cite{FRS1,FRS2} of von
Neumann algebras $F_{\rm red}(I)$, which are generated by
$\hat\pi(A(I))$ and the charged intertwiners $\psi^\sigma_e$ with
$\sigma$ localized in $I$. This net is relatively local with respect
to the subnet $\hat\pi(A)$, but non-local itself. The reduced field
net is covariant w.r.t.\ the unitary representation $\UU_{\hat\pi}$
implementing covariance of the observables. The operators
$\psi^\sigma_e$ satisfy braid-group commutation relations with
numerical coefficients (``$R$-matrices'') determined by the DHR
statistics operators $\varepsilon(\sigma_1,\sigma_2)$. They are
bounded operator versions of the chiral exchange fields discussed in
the Introduction.

If $u$ is a charge transporter $u:\sigma \to\hat\sigma$ with
$\hat\sigma$ localized in $\hat I$, then
\begin{equation}
\hat\pi(u)\psi^\sigma_e=\psi^{\hat\sigma}_e 
\end{equation}
belongs to 
$F_{\rm red}(\hat I)$. As discussed in \cite{FRS2}, suitable
regularized limits of $\psi^\sigma_e$ as the localization of $\sigma$
shrinks to a point $y$, behave like point-like chiral ``exchange fields'',
generalizing $a(y)$, $b(y)$ and their adjoints displayed in (1.10). Since the
commutation relations survive in the limit, the latter satisfy
commutation relations with the same $R$-matrices as the former.  
Their correlations converge to (primary or descendant, depending on
the details of the limit chosen) conformal blocks, whose analytical
monodromy properties thus represent the $R$-matrices of the DHR
statistics. 

The reduced field net does not comply with
the axioms for a (non-local) chiral extension of $A$ (in the sense of
\cite{LR} or \cite{ALR}) because its local algebras are not factors
(and as a consequence, the vacuum vector in $\HH_0$ is a cyclic, but not 
a separating vector). However, every (non-local) field extension $B$ on
a Hilbert space $\HH_B$ can be ``embedded'' in (an amplification of)
the reduced field net as follows \cite{LR}. Let $(\gamma,v,w)$ be the
Q-system associated with the inclusion $\pi(A)\subset B$, and
$(\theta,w,x)$ be the dual Q-system. $\theta$ is a DHR endomorphism of
$A$; we may write 
\footnote{In the sequel, indices
  $s,t,\ldots$ label the irreducible DHR sectors, while $p,q,\ldots$
  label the irreducible subrepresentations of $\pi$ which may come
  with multiplicities: $\pi\simeq\bigoplus\pi_p=\bigoplus
  n^s\pi_s(p)$. Indices $i,j,\ldots$ will label the irreducible
  components of $\Theta_+$ as in Sect.\ 4.}  
\begin{equation}
[\theta]=\bigoplus_{s:\rm inequivalent\;irreducibles} n^s[s]
=\bigoplus_{p:\rm irreducibles}[s(p)].
\end{equation}
Consequently $x:\theta\to\theta^2$ has an expansion of the form
\begin{equation}
x=\sum_{p,q,r;e} \lambda_{pq}^r(e)\cdot w^p\rho_p(w^q)\;T_e\;w^r{}^*
\end{equation}
where $\rho_p\equiv\rho_{s(p)}$ are the representatives of the sectors
$[s(p)]$ as before, $w^p:\rho_p\to\theta$ form a complete system of
orthonormal isometries in $A$, and $T_e\in A$ form orthonormal bases of
$\Hom(\rho_r,\rho_p\rho_q)$. The numerical coefficients
$\lambda_{pq}^r(e)\in \CC$ are ``generalized Clebsch-Gordon
coefficients'' characteristic for the inclusion $\pi(A(I))\subset
B(I)$. Then, the charged isometry $v\in B$ can be represented in terms
of operators from $F_{\rm red}$ as 
\begin{equation} 
v=\sum_{p,q,r;e} \lambda_{pq}^r(e)\cdot
E^p\hat\pi(w^q)\;\psi^{\rho_q}_e\;E^r{}^* =\sum_{p,q,r;e}
\lambda_{pq}^r(e)\cdot \pi(w^q)E^p\;\psi^{\rho_q}_e\;E^r{}^* 
\end{equation}
where $E^p:\hat\HH\to\HH_B$ are the partial isometries which identify
the irreducible subrepresentation $\HH_{s(p)}\subset\hat\HH$ with the
irreducible subrepresentation $\HH_p\subset\HH_B$, and are zero on the
complement. It follows that the charged intertwiners
$\psi_q:=d(\theta)^{\frac12}\cdot \pi(w^{q*})v$ of the chiral
extension $B$ (cf.\ App.\ A) arise as the characteristic linear
combinations  
\begin{equation}
\psi_q=d(\theta)^{\frac12}\sum_{p,r ;e}
\lambda_{pq}^r(e)\cdot E^p\;\psi^{\rho_q}_e\;E^r{}^*
\end{equation}
of charged intertwiners from $F_{\rm red}$ (possibly amplified by
multiplicities of sectors $[s]$ in $\HH_B$). The algebras $B(I)$
generated by these linear combinations do have the vacuum as a cyclic
{\em and} separating vector. Remarkably, in case $B$ is local
(or graded local), then the specific linear combinations (5.7) satisfy
(graded) local commutativity, although the individual summands
$\psi^{\rho}_e$ also in this case satisfy proper braid group
commutation relations.  
 \vskip3mm
{\bf 5.2. Application to BCFT.}
\vskip2mm
After these preliminaries, we return to boundary CFT. We formulate the
main result of this section:

{\bf 5.1 Proposition:} Let $\sigma$, $\bar\tau$ be irreducible DHR
endomorphisms, localized in $I$ and $J$, respectively, such that
$\sigma\bar\tau\prec\Theta_+$. Then the charged intertwiners
$\psi_i\in B\ind_+(O)$, $i=1\dots Z_{[\sigma][\tau]}$, for the inclusion 
$\pi(A_+(O))\subset B\ind_+(O)$ can be represented as 
\begin{equation}
\psi_i=\sum_{p,q;g,h} \varphi_{q,i}^p(g,h)\cdot
E^q\;\psi^\sigma_g\,\psi^{\bar\tau}_h\;E^p{}^* 
\end{equation}
with numerical coefficients $\varphi_{q,i}^p(g,h)$ to be specified in
Cor.\ 5.2 below. Here, the sums over $p$ and $q$ extend over the
irreducible subrepresentations of $\pi$, $h$ and $g$ stand for
orthonormal bases of intertwiners $T_h:\rho_{s(p)}\to\rho_t\bar\tau$ and
$T_g:\rho_t\to\rho_{s(q)}\sigma$, respectively, and sum over the
intermediate sectors $[t]$ is implicit in the sum over the
``channels'' $g$ and $h$.

{\it Proof:} Let us first consider the case of the reference
double-cone $O=I\times J=(y,z)\times(-z,-y)$ as discussed in Sect.\
4. We recall from (4.15) that the operators
$\psi_i=\varphi(t_i)=t_i\pi(R_{\tau})\in B\ind_+(O)\subset B$ are
intertwiners $\psi_i:\id_B\to\alpha_\sigma^-\alpha_{\bar\tau}^+$. 
Equivalently (because $A$ and $v$ generate $B$), they satisfy 
\begin{equation}
\psi_i\pi(a)=\pi(\sigma\bar\tau(a))\psi_i \qquad (a\in A)
\end{equation}
and 
\begin{equation}
\psi_iv=\alpha_\sigma^-\alpha_{\bar\tau}^+(v)\psi_i= 
\pi[\sigma(\eps(\theta,\bar\tau))\eps(\sigma,\theta)^*]v\psi_i.
\end{equation}
Now let $\UU:\pi\to\theta$ implement the unitary equivalence 
between the representation $\pi$ of $A$ on $\HH_B$ and the
representation through the DHR endomorphism $\theta$ on $\HH_0$, and
let $\varphi_i:=\Ad_\UU(\psi_i)$. Under $\Ad_\UU$, (5.9) and (5.10) 
translate into
\begin{equation}
\varphi_i\theta(a)=\theta\sigma\bar\tau(a)\varphi_i,
\end{equation}
i.e., $\varphi_i\in\Hom(\theta,\theta\sigma\bar\tau)\subset A$, and in
addition the linear condition on $\varphi_i$
\begin{equation}
\varphi_i\,x=
\theta[\sigma(\eps(\theta,\bar\tau))\eps(\sigma,\theta)^*]x\,\varphi_i,
\end{equation}
because $\Ad_\UU(v)=x$ \cite{LR}. Finally, the normalization (4.16)
of $\psi_i$ turns into the normalization
\begin{equation}
\varphi_i^*\varphi_j=d(\sigma)d(\tau)\cdot\delta_{ij}.
\end{equation}

Introducing a basis of the space $\Hom(\theta,\theta\sigma\bar\tau)$,  
we conclude:

{\bf 5.2 Corollary:} Consider the finite linear problem (5.12) to be
solved within the DHR category of $A$, i.e., $\varphi_i\in
\Hom(\theta,\theta\sigma\bar\tau)$. Let $\varphi_i$ be its solutions,
subject to the normalization (5.13). They have an expansion
\begin{equation}
\varphi_i=\sum_{p,q;g,h} \varphi_{q,i}^p(g,h)\cdot
w^qT_gT_hw^p{}^* 
\end{equation}
where $w^p=\UU^*E^p\rest_{\HH_{s(p)}}:\rho_p\to\theta$ are orthonormal
isometries. Transformed back to $\HH_B$, we have (5.8), where the
numerical coefficients $\varphi_{q,i}^p(g,h)$ are given by
\begin{equation}
\varphi_{q,i}^p(g,h)= T_h^*T_g^*w^q{}^*\varphi_iw^p\in\Hom(\rho_p,\rho_p)=\CC.
\end{equation}

This concludes the proof of Prop.\ 5.1 in the case of the reference
double-cone $O$. Now, we may change the localization to any other
double-cone $\hat O=\hat I\times \hat J$. Similar as in (5.3), we
multiply $\psi_i$ from the left with the charge transporter
$\pi(U_\sigma\sigma(U_{\bar\tau}))$ where
$U_\sigma:\sigma\to\hat\sigma$ and
$U_{\bar\tau}:{\bar\tau}\to\hat{\bar\tau}$ with the desired
localizations. From $\pi(\sigma(U_{\bar\tau}))E^q\psi^\sigma_g= 
E^q\hat\pi(\sigma(U_{\bar\tau}))\psi^\sigma_g=
E^q\psi^\sigma_g\hat\pi(U_{\bar\tau})$ and (5.3), we conclude that
(5.8) in fact holds for the charged intertwiners associated with
arbitrary double-cones $\hat O$, substituting only $\hat\sigma$ for
$\sigma$ and $\hat{\bar\tau}$ for $\bar\tau$. Dropping the $\,\hat{}\,$
symbols, we may equally well assert that the structure (5.8) holds for
any double-cone. 

This completes the proof of Prop.\ 5.1 in the general case. \QED

We note that eq.\ (5.12) is quite similar to the condition
Def.\ 5.5 in \cite{FRS}. 

Note that $\psi^{\sigma}_g$ belongs to $F_{\rm red}(I)$, and 
$\psi^{\bar\tau}_h$ belongs to $F_{\rm red}(J)$. We have thus
geometrically separated the ``left'' and ``right'' charges of the 
charged intertwiners, by representing them as linear combinations of  
{\em bilocalized products} of charged intertwiners from $F_{\rm red}(I)$ 
and $F_{\rm red}(I)$. The specific coefficients
$\varphi_{q,i}^p(g,h)$, arising through the solution of a linear
problem in the DHR category involving the dual canonical endomorphism
$\theta$ (Cor.\ 5.2), are algebraic invariants for the (non-local)
chiral extension $\pi(A)\subset B$.   

Assuming the same regularity of the point-like limits $\hat
O\to(t,x)$ as in \cite{FRS2}, we infer the convergence of
$n$-point correlations of $\psi_i(t,x)$ to characteristic linear
combinations of $2n$-point conformal blocks involving the arguments
$t+x$ and $t-x$. (Clearly, the limit cannot be effectuated by
the action of the M\"obius group. Instead, one has to use the local
implementers which implement the local action of the M\"obius group on
$F_{\rm red}(I)$ and act trivially on $F_{\rm red}(J)$, and vice
versa, to obtain a local action of $PSL(2,\RR) \times PSL(2,\RR)$ on 
$F_{\rm red}(I)\vee F_{\rm red}(J)$ and hence on $B(O)$.)

The coefficients $\varphi_{b,i}^a(g,h)$ are affected neither by the
transport from $O$ to $\hat O$ nor (up to some overall normalization)
by the point-like limit. We conclude that Cardy's observation,
originally derived from Ward identities in minimal models, is in fact
a model-independent feature of boundary CFT, reflecting purely
algebraic structures of the associated (non-local) chiral extension
$\pi(A)\subset B$. The relative coefficients of the representation of
local $n$-point correlation functions as linear combinations of
$2n$-point conformal blocks are the products of $n$ coefficients
$\varphi_{q,i}^p(g,h)$ according to the contributing channels.

It should be remarked that, according to the structure (5.8), while
the initial and final sectors of $\psi_i$ necessarily belong to
$\HH_B$, the intermediate sectors $[t]$ may range over all DHR sectors
of the chiral net $A$, as can be nicely seen in the examples (1.11)
and (1.12). The correlation functions of boundary CFT therefore
carry information also about those chiral sectors which are not
present in the Hilbert space of its local fields. 

\section{Varying the boundary conditions}
\setcounter{equation}{0} 

As we have seen, the chiral extension $I\mapsto B(I)$ of $I\mapsto
A(I)$ determines not only the Hilbert space $\HH_B\simeq\bigoplus_s
n^s\HH_s$ of the boundary CFT, but also the detailed charge
structure of its local fields as in (5.8) and, as a consequence
illustrated by the example (1.8), (1.9), the behavior of the local
fields and their correlations close to the boundary $x=0$. 

In this section, we want to {\em vary the boundary conditions} by
varying the (non-local) chiral extension $B$. As is well known from
\cite{BEK1}, there is a finite system of inequivalent (non-local)
chiral extensions $B_a$ which all give rise to the same coupling
matrix $Z_{[\sigma][\tau]}$. In the language of modular categories
\cite{FS1,LRo,KO}, these extensions correspond to Morita equivalent
Frobenius algebras \cite{FRS,Os}. We want to show here, that they even give
rise to boundary CFT's with locally isomorphic subfactors
$A_+(O)\subset B\ind_{a,+}(O)$. In view of Thm.\ 4.1, this means that
they all share the local structure of the same Minkowski space CFT
$B^\alpha_2$. 

Our result is essentially a corollary to a result in \cite{BE3} making
use of \cite{BEK1}.  

Let $\iota:A(I)\to B(I)$ be the inclusion homomorphism for a given
(non-local) chiral extension $\pi(A)\subset B$, and consider the
system $\XX=\{a:A(I)\to B(I)\}$ of inequivalent irreducible
subhomomorphisms of $\iota\circ\rho$ as $\rho$ ranges over the DHR
endomorphisms of $A$ localized in $I$.\footnote{In
  the terminology of categories \cite{FS2,KO,MM2} (where a Q-system is
  a Frobenius algebra), $a\in \XX$ are the irreducible modules of the
  Frobenius algebra, cf.\ e.g., \cite[Lemma 5.24 and Chap.\ 6]{FS2},
  forming the objects of the module category. If $B$ is local, the
  Frobenius algebra is commutative. In this case, Kirillov and Ostrik
  have shown \cite{KO} that the module category is again a monoidal
  (tensor) category. In the same situation, B\"ockenhauer and Evans
  have found \cite[Thm.\ 3.9]{BE1} a bijection between the elements
  $a\prec\iota\rho$ of $\XX$ and the irreducible subendomorphisms
  $\beta\prec\alpha^\pm_\rho$ of $\alpha$-induced endomorphisms
  ($\dim\Hom(\iota\rho,\iota\sigma) =
  \dim\Hom(\alpha^\pm_\rho,\alpha^\pm_\sigma)$). These two results
  relate to each other in such a way that the monoidal product
  $a_1\times a_2$ coincides with the composition of endomorphisms
  $\beta_1\circ\beta_2$. In the non-local case, there is no such bijection.)}

Each $a\in\XX$ naturally gives rise to a Q-system $(\theta_a,w_a,x_a)$
(where $\theta_a=\bar a a \prec \bar\rho\,\theta\rho$ is a DHR
endomorphism because $a\prec \iota\rho$) and hence defines an inclusion  
$A\subset B_a$ as in \cite{E,BE3} (``varying the iota vertex''), i.e.,
(non-local) chiral extensions $\pi_a(A)\subset B_a$ with inclusion
homomorphisms $\iota_a$ such that 
$\theta_a=\bar \iota_a\iota_a$. We may call the family $B_a$ 
(as $a$ varies over $\XX$) the {\it DHR orbit} of the given
extension $B$. (Warning: The association $a\mapsto B_a$ is in
general not injective, see below.)   

Each member of the DHR orbit induces a boundary CFT $B\ind_{a,+}$ as
well as a Minkowski space CFT $B^\alpha_{a,2}$ by the
$\alpha$-induction construction. Although the associated
representations $\pi_a\simeq\theta_a$ in general differ from each
other, the following local isomorphy holds.

{\bf 6.1 Proposition:} The inclusions $A_+(O)\subset B\ind_{a,+}(O)$ are
isomorphic throughout the DHR orbit. The same holds true for the
inclusions $A_2(O)\subset B^\alpha_{a,2}(O)$.  
 
{\it Proof:} We recall from Sect.\ 4, that the algebraic structure of
the subfactors of interest is coded in the numerical coefficients 
$\zeta_{ij,ef}^k$ of their Q-systems. The latter, in turn, arise as
expansion coefficients (4.19) of the monoidal product
$t_i\alpha^+_{\tau_i}(t_j)$ of intertwiners
$t:\alpha^+_\tau\to\alpha^-_\sigma$ between $\alpha$-induced
endomorphisms of both signs. In \cite[p.\ 21]{BE3}, a bijection
$\beta_a$ between the intertwiner spaces
$\Hom(\alpha^+_{\tau},\alpha^-_{\sigma})$ and 
$\Hom(\alpha^+_{a,\tau},\alpha^-_{a,\sigma})$ for $\alpha$-inductions
to the extensions $B_a$ within a DHR orbit was established. It is
therefore sufficient to show that this bijection respects the monoidal
product.   

Let $M=B(I)$ and $N=A(I)$. For $a\in\XX$, let $\bar a:M\to N$ be a
conjugate homomorphism, and  $\bar a(M)\subset N\subset M_a$ the Jones
basic construction \cite{J} associated with the subfactor $\bar
a(M)\subset N$. Let $\iota_a:N\to M_a$ be the inclusion homomorphism
of $N$ into $M_a$, and $\bar\iota_a:M_a\to N$ a conjugate homomorphism
such that $\bar\iota_a(M_a)=\bar a(M)\subset N$. Then $\varphi_a=\bar
a\inv\circ\bar\iota_a:M_a\to M$ is an isomorphism. 

Now, if $\rho$ and $\bar\rho$ are conjugate DHR endomorphisms of $A$
localized in $I$ and $a\prec\iota\rho\vert_N$, then
$\theta_a=\bar\iota_a\iota_a=\bar aa$ is contained in (the restriction
to $N$ of) $\bar\rho \bar\iota\iota \rho =\bar\rho\,\theta^I\rho$
which is again a DHR endomorphism localized in $I$. The statistics
operators $\eps^\pm(\tau,\theta_a)$ enter the definition of
$\alpha$-induction $\alpha^\pm_{a,\tau}$, cf.\ App.\ B.
According to \cite[p.\ 455f]{BEK1}, if $T\in\Hom(a,\iota\rho)\subset
M$ is isometric, then 
\begin{equation}
U^\pm_\tau =T^*
\iota(\eps^\pm(\tau,\rho))\alpha^\pm_\tau(T) \in
\Hom(\alpha^\pm_\tau a,a\tau)\subset M
\end{equation} 
is unitary, and $\eps^\pm(\tau,\theta_a)=\bar
a(U^\pm_\tau)\eps^\pm(\tau,\bar a\iota)$. One finds 
\begin{equation}
\alpha^\pm_{a,\tau}=\varphi_a\inv\circ
\Ad_{U^\pm_\tau}\circ\alpha^\pm_\tau\circ\varphi.
\end{equation}
Consequently, the bijection $\beta_a:\Hom(\alpha^+_{\tau},\alpha^-_{\sigma})\to
\Hom(\alpha^+_{a,\tau}\alpha^-_{a,\sigma})$ is given by 
\begin{equation}
\beta_a(t)=\varphi_a\inv(U^-_\sigma t U^+_\tau{}^*).
\end{equation}

In order to show that $\beta_a$ respects the monoidal product, we have
to show that 
\begin{equation}
\varphi_a\inv\big(U^-_{\sigma_1\sigma_2}t_1\cdot\alpha^+_{\tau_1}(t_2)
U^+_{\tau_1\tau_2}{}^*\big)=\varphi_a\inv\big(U^-_{\sigma_1}t_1
U^+_{\tau_1}{}^*\big)\cdot\alpha^+_{a,\tau_1} 
\varphi_a\inv\big(U^-_{\sigma_2}t_2U^+_{\tau_2}{}^*\big)
\end{equation}
which due to (6.2) is equivalent to
\begin{equation}
U^-_{\sigma_1\sigma_2}t_1\cdot\alpha^+_{\tau_1}(t_2)
U^+_{\tau_1\tau_2}{}^*= U^-_{\sigma_1}t_1
U^+_{\tau_1}{}^*\cdot
U^+_{\tau_1}\alpha^+_{\tau_1}(U^-_{\sigma_2}t_2U^+_{\tau_2}{}^*)
U^+_{\tau_1}{}^*. 
\end{equation}
Using ``naturality'' of the DHR braiding with respect to
$\alpha$-induction, as expressed, e.g., in \cite[Eq.\ (14)]{BEK1}, we
find 
\begin{equation}
U^+_{\tau_1}\alpha^+_{\tau_1}(U^+_{\tau_2})=U^+_{\tau_1\tau_2}
\end{equation}
and similar for $U^-_\sigma$, which implies (6.5). This completes the
proof. 
\QED

Since the structure of the local subfactors in the case of Minkowski
space extensions $B_2$ of $A\otimes A$ determines the global structure
(thanks to the ``unbroken symmetry'', i.e., existence of a global
conditional expectation in this case, cf.\ Sect.\ 2), the associated
two-dimensional theories $B^\alpha_{a,2}$ may in fact be considered as
identical. 

In contrast, the boundary CFT nets $B\ind_{a,+}$ are defined on
different Hilbert spaces $\HH_a$ given by $\pi_a\simeq \theta_a=\bar\iota_a\iota_a$. 
In particular, in spite of the algebraic isomorphism of the local
subfactors, the corresponding bi-localized charge structures as in Prop.\
5.1 differ among different members within the DHR orbit. As a
consequence, exemplified by the example (1.11) and (1.12), also the
scaling behavior of the local fields towards the boundary differs.

The DHR orbit associates several BCFT's to a given one. E.g.,
the ``Cardy case'' discussed in the literature \cite{C,FFFS,Z} is the
DHR orbit of the trivial extension $B=A$, 
$\iota=\id$,  which includes $B_+=A\dual_+$. The elements of $\XX$
in this case are labelled by the sectors $\iota_\rho\equiv\rho$ of
$A$. To be more specific, the Hilbert spaces $\HH_{\rho}$
carry the representation
$\pi_\rho\simeq\theta_\rho\equiv\bar\rho\rho$ of $A$ 
and hence of $A_+$ and $A\dual_+$. Thus, in the Cardy case, the
members of the DHR family are just the extensions
$\pi_\rho(A_+)\simeq\bar\rho\rho(A_+)\subset\bar\rho\rho(A\dual_+)$.
The non-trivial charge structure of the ``charged fields'' of
$A\dual_+$ arises through the non-trivial action of
$\bar\rho\rho$ on the charge transporters $u:\sigma^I\to\sigma^J$
(cf.\ Remark 3 after Def.\ 2.1). In the Ising model, there are three
sectors $[0]$, $[\frac1{16}]$, and $[\frac12]$. The corresponding
chiral extensions $B_s$ are, in turn, $A$ itself, $CAR$ (cf.\ Sect.\
2), and again $A$ itself (exemplifying the non-injectivity of the 
association $a\mapsto B_a$). The boundary field nets $B\ind_{s,+}$ are
generated by $A_+$ and, in turn, charged intertwiners of the structure
$\phi_0$ as in (1.11), $\phi_1$ as in (1.12), and again $\phi_0$. In fact,
$B\ind_{0,+}$ and $B\ind_{\frac12,+}$ both coincide with $A\dual_+$. A
more refined structure distinguishing between $0$ and $\frac12$ will
be discussed in the next section.  

The main problem, however, is the classification of the other orbits,
if there are any. By the results of Sect.\ 2, this amounts to the
classification of non-local chiral extensions $\pi(A)\subset B$,
reformulated according to App.\ A as the classification of Q-systems
in the DHR category of superselection sectors of $A$.\footnote{See
  also \cite[Thm.\ 1]{Os} according to which every module category
  arises as the module category of some Frobenius algebra.} From
As explained in Sect.\ 3.2, this is a finite-dimensional problem and
it has only finitely many solutions. Of course, complete
classifications can be expected only when the chiral observables $A$
are specified, see e.g., \cite{KL}.    

We speculate that each DHR orbit of non-local chiral extensions 
$A\subset B_a$ contains 
a distinguished element which is local, at least if the coupling
matrix $Z_{[\sigma][\tau]}$ is of type I \cite{BEK1}. The argument
could go like this. Every element of the DHR orbit defines the same theory 
$B_2$ on Minkowski space by the $\alpha$-induction prescription, see
above. This theory in turn has a pair of maximal chiral subalgebras
$A^{\rm max}_L\supset \pi^{\rm max}_L(A)$ and $A^{\rm max}_R\supset
\pi^{\rm max}_R(A)$, where $\pi^{\rm max}_L$ and $\pi^{\rm max}_R$ are 
determined by the ``vacuum block'' of the coupling matrix
$Z_{[\sigma][\tau]}$. (We expect that these coincide with $\hat A_L$
and $\hat A_R$ mentioned in the end of Sect.\ 4.) If $Z$ is of type 1,
$\pi^{\rm max}_L$ and $\pi^{\rm max}_R$ are equivalent, and we may
suppress the subscript. We conjecture that the local chiral extension
$\pi^{\rm max}(A)\subset A^{\rm max}$ is a distinguished {\em local}
element of the orbit. Thus, classification of DHR orbits of
BCFT would be reduced to classification of local chiral extensions,
cf.\ \cite{KL,LR}, or of commutative Frobenius algebras \cite{KO}. We
hope to return to this conjecture in a separate work.   

\section{Partition functions and modularity}
\setcounter{equation}{0} 
We mention in this section aspects of modular invariant partition
functions, as far as they can be easily derived in our framework. 
Let us recall, however (cf.\ Sect.\ 1), that in our
approach Modular Invariance of the partition function is not a first
principle. Therefore, the natural appearance of the matrix $Z$ (both
as the coupling matrix of left and right chiral sectors in the
Minkowski space theory $B^\alpha_2$ and as the coupling matrix for the 
bi-localized charge structure, cf.\ Prop.\ 4.4), its automatic modular
invariance \cite{BEK1}, and the validity of relations (7.4) and (7.7)
below also in the flat space QFT framework, is a remarkable fact about the
intrinsic structure of Minkowski space CFT with or without boundary.

The structure of the system $\XX=\{a\prec
\iota\rho \;\hbox{irreducible}\}$ associated with a BCFT (cf.\ Sect.\
6) defines a
``nimrep'' (non-negative integer matrix representation) of the fusion
rules $[s][t]=\bigoplus_u N^{st}_u[u]$ 
of the superselection sectors. Namely, if $a$ belongs to $\XX$, then
$a\prec \iota\rho$ for some DHR endomorphism $\rho$, hence then
$a\rho_s\prec\iota\rho\rho_s$, and every irreducible component of
$a\rho_t$ again belongs to $\XX$. Hence 
 \begin{equation}
[a\rho_s]=\bigoplus_{b\in\XX} n^s_{ab} [b] \qquad\hbox{with}\qquad\sum_b 
n^s_{ab} n^t_{bc} = \sum_u N^{st}_u n^u_{ac}.
\end{equation}
This implies that the diagonal matrix elements $n^s_{aa}$ are the
multiplicities of $\rho^s$ within $\theta_a=\bar a a$, thus
\begin{equation}
\HH_a =\HH_{B_a} = \bigoplus_s n^s_{aa}\HH_s \equiv \bigoplus_s n^s_{aa}\HH_s. 
\end{equation}

In the literature on boundary CFT in Statistical Mechanics (for a
review see, e.g., \cite{Z}) one discusses also theories defined on
Hilbert spaces 
\begin{equation}
\HH_{ab}=\bigoplus_s n^s_{ab}\HH_s.
\end{equation}
This leads us to consider ``non-diagonal'' boundary CFT nets $O\mapsto
B_{ab,+(O)}\supset\pi_{ab}(A_+(O))$ and the associated (non-local)
chiral nets $I\mapsto B_{ab}(I)\supset \pi_{ab}(A(I))$ which are
defined on $\HH_{ab}$ carrying the DHR representation
$\pi_{ab}\simeq\bar ab$, for any pair $a,b\in\XX$.  
These theories arise through the reducible subfactor associated with
$\theta=(\bar a\oplus \bar b)\circ(a\oplus b)$. 
If $a\neq b$, the Hilbert spaces $H_{ab}$ do not
contain the vacuum vector (because $n^0_{ab}=\delta_{ab}$), so that the
standard theory of chiral extensions as applied in Sect.\ 2--6 cannot
be used. We expect nevertheless (without elaborating) that the
results of the previous sections largely carry over to these theories
as well, and allow to make precise contact with the Statistical
Mechanics interpretation along the following lines. 

The partition function for the spectrum of the chiral conformal
Hamiltonian of the boundary CFT on $\HH_{ab}=\bigoplus_s n^s_{ab} \HH_s$ is 
\begin{equation}
Z_{ab}(\beta)=\Tr_{\HH_{ab}}
\pi_{ab}(\exp-\beta (L_0-{\textstyle\frac c{24}})) = \sum_sn^s_{ab}\,
\chi_s(\beta).
\end{equation}
$n^s$ being a nimrep of the (commutative) fusion rules, its joint
spectrum is given by the matrix elements of the modular matrix $S$
(note that in the algebraic approach, complete rationality implies
non-degeneracy of the braiding \cite{KLM}, and hence the DHR
statistics defines a unitary representation of the modular group
$SL(2,\ZZ)$ \cite{FRS2}), i.e., one has the ``Cardy equation''
\begin{equation}
n^s_{ab} = \sum_{t}\psi_{at} \frac{S_{st}}{S_{0t}}\psi_{bt}^*.
\end{equation}
Inserting this expansion in the partition function $Z_{ab}$, and
taking for granted the modular transformation law of the chiral 
characters $\chi_s(\beta)$, one obtains 
\begin{equation}
Z_{ab}(\beta)= \sum_{t}\psi_{bt}^*\chi_t(\hat\beta)\psi_{at},
\end{equation}
where $\hat\beta=4\pi^2/\beta$ is the modular transform of the
inverse temperature $\beta$.  Usually \cite{Z}, the right-hand 
side of this formula is reinterpreted as a matrix element of the
conformal Hamiltonian of the Minkowski space theory between a pair of 
so-called ``Ishibashi boundary states'' $\vert a \rangle = 
\sum_t\psi_{at}\vert t\rangle$, which weakly realize the boundary
condition $T_L=T_R$: 
\begin{equation}
Z_{ab}(\beta)=\langle b\vert\exp-{\textstyle \frac12}
\hat\beta(L_{L0}+L_{R0}-{\textstyle\frac c{12}})\vert a\rangle.
\end{equation}
These Ishibashi states, however, are linear combinations of 
non-normalizable vectors in the Hilbert spaces $\HH_t\otimes \HH_t$. It was
pointed out, e.g., in \cite{HS} that Ishibashi states, rather than vector
states on $A\otimes A$, should be considered as KMS (= Gibbs in
this case) states on $A$, where the second copy of $A$ appears via 
Tomita's Modular Theory \cite[Chap.\ VI, Thm.\ 1.19]{T} as the
commutant of $A$ in the GNS representation of the KMS state. 

While we have not elaborated these issues, we hope to arrive, in a
future publication, at a better algebraic understanding of the
structures outlined in this section.

\section{Conclusion}
\setcounter{equation}{0} 
We have classified boundary conformal quantum field theories in terms
of chiral extensions of the underlying local chiral observables $A$. 
These extensions, which are in general non-local, are in turn
classified in terms of Q-systems (Frobenius algebras) in the DHR modular
category of superselection sectors of $A$. We have analysed how
general structural properties of the chiral observables are
transmitted to the local algebras of the BCFT. Among other things, we
have shown the absence of DHR superselection sectors of the latter
(Sect.\ 2).

A chiral extension determines both a BCFT and a Minkowski space CFT. 
Well away from the boundary, these two theories are algebraically
indistinguishable (Sect.\ 4). Only near the boundary, the breakdown of
symmetry changes the algebraic structure. This effect is exhibited in
the bi-localized charge structure of the local fields in the BCFT.
This structure can be derived (and explicitly computed) from the
superselection structure  (the DHR modular category) 
of the chiral observables. It may be regarded
as an algebraic invariant for the embedding of the latter into the
full theory (Sect.\ 5). The bi-localized charge structure in turn
determines the scaling behavior of the local fields with $x\to 0$. In
this sense, the boundary ``conditions'' on the non-chiral fields of
BCFT are in fact rather a derived feature.

BCFT's associated with the same chiral  observables can be grouped into
families (``DHR orbits'') which are algebraically isomorphic well away
from the boundary, but differ near the boundary. The members
of each orbit may thus be interpreted as the different ways a
Minkowski space CFT may ``react'' to the presence of a boundary; but
it can (in general) not be considered as different representations of
the same abstract theory on the half-space (Sect.\ 6).

Each DHR orbit is accompanied by a ``nimrep'' of the fusion rules of
the chiral observables, which controls the modular behavior of the
partition function of the conformal Hamiltonian (Sect.\ 7). 

One may also study boundary CFT with two boundaries \cite{Z}, 
corresponding to a QFT on a strip $0<x<L$. The formulae (1.2), (1.4), (1.16)
etc.\ for the chiral observables pertain to that situation as well,
provided $t\pm x$ are interpreted as angular coordinates of the
circle, adjusted with a normalization factor $L/2\pi$, rather than
cartesian coordinates of the lightlike axes. We refrain in this
article from elaboration on local extensions of the chiral observables
on the strip. 

\appendix
\section{Q-systems and algebras of charged intertwiners}
\setcounter{equation}{0}

We give a brief reminder of the notion of {\it Q-system}
associated with a subfactor $N\subset M$ of type $I\!I\!I$ von Neumann
algebras, and then present a Lemma concerning the generation of $M$ in
terms of charged intertwiners. This lemma is the obvious
generalization of an argument used in \cite{KLM} in a special case.

A subfactor $N\subset M$ is irreducible if $N'\cap M=\CC\cdot\Eins$. 
The {\em index} $[M:N]$ is the optimal bound $\lambda\geq 1$ such that
there is a conditional expectation ${\cal E}:M\to N$ satisfying the lower
operator bound ${\cal E}(m^*m)\geq \lambda\inv\cdot m^*m$. The 
{\em dimension} $d(\rho)$ of an endomorphism $\rho\in\End(N)$ is the
square root of the index $[N:\rho(N)]$.  

The condition of finite index is equivalent to the property that, with
$\iota:N\to M$ the inclusion homomorphism, there is a ``conjugate''
homomorphism $\bar\iota:M\to N$ and a ``canonical'' pair of isometric 
intertwiners $v:\id_M\to\gamma:=\iota\bar\iota\in\End(M)$ in $M$ and
$w:\id_N\to\theta:=\bar\iota\iota\in\End(N)$ in $N$, such that
$\iota(w)^*v=\lambda^{-\frac12}\Eins_M$ and  
$\bar\iota(v)^*w=\lambda^{-\frac12}\Eins_N$. Then,
${\cal E}(m)=\iota(w)^*\gamma(m)\iota(w)$ is the (unique, if $N\subset M$ 
is irreducible) conditional expectation. $\gamma$ and $\theta$ are the
``canonical'' and ``dual canonical'' endomorphisms associated with the
subfactor, and $d(\gamma)=d(\theta)=\lambda=[M:N]$.

A {\it Q-system} in $M$ is a triple $(\rho,T,S)$ where $\rho\in\End(M)$
is an endomorphism of $M$, and $T$ and $S$ are isometric intertwiners
$T:\id\to\rho$ and $S:\rho\to\rho^2$ in $M$, satisfying the
relations \footnote{In more general
  frameworks, such as Frobenius algebras in tensor categories
  \cite{FS2}, one has to require in addition an equivalent of
  $SS^*=\rho(S^*)S$; in a C* context as ours, the latter relation
  follows from the remaining ones \cite[Sect.\ 6]{LRo}.} 
\begin{equation} 
T^*S = \rho(T^*)S = \lambda^{-\frac12}\cdot \Eins, \qquad SS=\rho(S)S
\end{equation}
A Q-system in $M$ determines a subfactor $N\subset M$ of index
$\lambda$ in terms of data of $M$ as the image $N:={\cal E}(M)$ of the
conditional expectation ${\cal E}: M\to N$, defined by ${\cal E}(m) :=
T^*\rho(m)T$. Thus, the Q-system for $N\subset M$ is
$(\gamma,v,\iota(w))$. Likewise, the Q-system in $N$ for
$\bar\iota(M)\subset N$ (the {\em dual Q-system for $N\subset M$}) is
$(\theta,w,\bar\iota(v))$.  

By Jones' ``basic construction'' \cite{J}, a subfactor $N\subset M$
determines (up to unitary equivalence) another subfactor $M\subset
M_1$, isomorphic to $\bar\iota(M)\subset N$. The basic construction
applied to $\bar\iota(M)\subset N$, recovers $N\subset M$ (up to
isomorphism), hence $N\subset M$ is also determined by its {\em dual}
Q-system $(\theta,w,x)$ in $N$. 

Given a Q-system $(\theta,w,x)$ in $N$, a {\em concrete} realization
of $M$ results if one finds a representation of $N$ in a Hilbert space 
$\HH$ and an isometry $v$ in $\BB(\HH)$ such that $(\gamma,v,w)$ form
a Q-system in $M:=N\vee \{v\}$ where $\gamma$ extends $\theta$ by setting
$\gamma(v):=x$. Then, with $\iota$ the inclusion map of $N$ into $M$
via its representation on $\HH$, $(\gamma,v,w)$ is the Q-system for
$N\subset M$, and $(\theta,w,x)$ is the dual Q-system. We make use of
this constructive scheme in Sect.\ 4.  

The conditions on $v$ which ensure that $(\gamma,v,w)$ form a
Q-system with $\gamma\rest_N=\theta$ and $\gamma(v)=x$, can be
formulated as an {\em algebra of charged intertwiners}, as follows. 

Let $N\subset M$ be a subfactor of finite index, and $(\gamma,v,w)$
and $(\theta,w,x)$ its Q-system and dual Q-system, and  
\begin{equation}
\theta(n) = \sum_i w^i\varrho_i(n)w^{i*} \qquad (n\in N)
\end{equation}
the decomposition of $\theta$ into irreducibles (choosing representatives
$\rho_i=\rho_j$ whenever $\rho_i$ and $\rho_j$ are equivalent), $\varrho_0
= \id$, $w^0 = w$. Then the {\em charged intertwiners}
\begin{equation}
\psi_i:=d(\theta)^{\frac12}\cdot w^{i*}v \in M
\end{equation} 
satisfy
\begin{equation}
\psi_i n=\varrho_i(n)\psi_i \qquad (n\in N);
\end{equation}
we say that ``$\psi_i$ carry charge $\varrho_i$''. The charged
intertwiner for $\varrho_0=\id$ is
\begin{equation}
\psi_0=\Eins,
\end{equation}
and whenever $\varrho_i=\varrho_j$, one has the normalization 
\begin{equation}
\psi_i^*\psi_j = d(\varrho_i)\cdot\delta_{ij}.
\end{equation} 
Together with $N$, the charged intertwiners generate $M$; more precisely,
every element of $M$ has a unique expansion 
\begin{equation}
m = \sum_k n_k\psi_k,\qquad (n_k={\cal E}(m\psi_k^*)\in N).
\end{equation}

As $x\in N$ is an intertwiner $x:\theta\to\theta^2$, it has a unique
expansion 
\begin{equation}
x = d(\theta)^{-\frac12}\sum_{ijk} w^{i}\varrho_i(w^j)\,\Gamma_{ij}^k\, w^{k*}
\end{equation}
with intertwiners $\Gamma_{ij}^k:\varrho_k\to\varrho_i\varrho_j$  in $N$. 
Transscribing the relations of the Q-system $vv=\gamma(v)v=xv$ and
$v^*=d(\theta)^{\frac12}w^*vv^* = d(\theta)^{\frac12}w^*\gamma(v^*)v = 
d(\theta)^{\frac12}w^*x^*v$ in terms of the charged intertwiners, one
arrives at 
\begin{equation}
\psi_i\psi_j= \sum_k \Gamma_{ij}^k\, \psi_k, \qquad 
\psi_i^* = \sum_j \Gamma_{ji}^{0*}\,\psi_j.
\end{equation}

We note that, by (A.6) alone, $\sum_it_i\psi_i=0$ with
 $N\ni t_i:\varrho_i\to\varrho$ implies $t_i=0$. Indeed, multiplying the sum
from the left by $\psi_j^*s^*$ ( $N\ni s:\varrho_j\to\rho$) one gets
$s^*t_j=0$. Since $s$ is arbitrary, $t_j=0$. Therefore, the relations
 (A.4--6) and (A.9) among the charged intertwiners impose constraints
on the ``coefficients'' $\Gamma_{ij}^k$, e.g.,
$\Gamma^k_{0j}=\delta_{jk}\Eins$, $\Gamma^k_{i0}=\delta_{ik}\Eins$, as
well as $\sum_n\Gamma^n_{ij} \Gamma^l_{nk}=
\sum_m\varrho_i(\Gamma^m_{jk})\Gamma^l_{im}$ (associativity of
the expansion (A.9)). Furthermore, $\Gamma_{ij}^0:\id\to\varrho_j\varrho_i$ 
vanish unless $\varrho_j$ is conjugate to $\varrho_i$, and $\sum_k
\Gamma^{k*}_{ij}\Gamma^{k}_{ij} d(\varrho_k)=d(\varrho_i)d(\varrho_j)$. 

The relation 
\begin{equation}
\sum_{ij}\Gamma_{ij}^{k*}\Gamma_{ij}^{k'}=d(\theta)\delta_{kk'}\cdot\Eins
\end{equation}
is of a different  status: it follows from $x^*x=1$, but 
seemingly not from the relations among the charged intertwiners alone.

This set of relations  (A.4--6) and (A.9) in $M$ is, like the
Q-system $(\gamma,v,w)$, a complete invariant for the subfactor
$N\subset M$, see Lemma A.2. 

{\bf A.1 Definition:} Let $\varrho_i$ be a finite system of pairwise
either inequivalent or equal irreducible endomorphisms of $N$ of
finite dimension among which $\varrho_0=\id$ occurs precisely once, and
$\Gamma^k_{ij}\in\Hom(\varrho_k,\varrho_i\varrho_j)\subset N$. 
An {\em algebra of charged intertwiners} for $N$ is a
system of operators $\psi_i$ satisfying (a) the 
intertwining property (A.4) for $\varrho_i$, (b) the normalizations as in
(A.5), (A.6), and (c) the algebra (A.9) with coefficients satisfying
(A.10), where $d(\theta):=\sum_id(\varrho_i)$.

The first statement of the following lemma
summarizes the above discussion; the second statement is the converse:
the algebra of charged intertwiners determines the subfactor and its
Q-system. While a special case underlies the argument leading to Cor.\
45 of \cite{KLM}, we think it appropriate to formulate the general case. 

{\bf A.2 Lemma:} (i) An irreducible subfactor $N\subset M$ determines
an algebra of charged intertwiners. In particular, the condition
(A.10) on the coefficients is automatic, if they arise in this way.

(ii) Let $\psi_i\in \BB(\HH)$ be an algebra of charged intertwiners
for $N$ with endomorphisms $\varrho_i\in\End(N)$ and coefficients
$\Gamma^k_{ij}\in N$, and $M$ be the algebra generated by $N$ and
$\psi_i$. Then the subfactor $N\subset M$ has Q-system $(\gamma,v,w)$ and dual
Q-system $(\theta,w,x)$, where (in turn) $\theta$ is 
defined as in (A.2) with the help of any complete orthogonal system of
isometries $w^i\in A$, $w:=w^0$, $x$ is defined as in (A.8),
$v:=d(\theta)^{-\frac12}\sum_iw^i\psi_i$, and by 
definition $\gamma$ extends $\theta$ by  $\gamma(nv):=\theta(n)x$. 

(iii) Two algebras of charged intertwiners with the same endomorphisms
in $\End(N)$ and coefficients in $N$ give rise to isomorphic 
subfactors (possibly on different Hilbert spaces), with the
isomorphism given by identification of the charged intertwiners. 

{\it Sketch of the proof:} (ii) It is straightforward to see that the
relations of the algebra of charged intertwiners ensure the following:  
$wn=\theta(n)w$, $x\theta(n)=\theta^2(n)x$, $vn=\theta(n)v$ ($n\in
N$); $w^*w=\Eins$, $v^*v=\Eins$, $x^*x=\Eins$;
$\theta(w^*)x=d(\theta)^{-\frac12}\Eins$,  
$w^*x=w^*\gamma(v)=d(\theta)^{-\frac12}\Eins$,
$w^*v=d(\theta)^{-\frac12}\Eins$; $vv=xv$, $xx=\theta(x)x$;
and $v^*=d(\theta)^{\frac12}\cdot w^*x^*v$. These include the defining
relations for $(\theta,w,x)$ to be a Q-system in $N$. The missing
information for $(\gamma,v,w)$ to be a Q-system is that $\gamma$ is an
endomorphism. It is also straightforward to see that $\gamma$ respects
products and the previous relations, and 
\begin{equation}
\gamma(v^*)=d(\theta)^{\frac12}\cdot \theta(w^*x^*)x = 
d(\theta)^{\frac12}\cdot \theta(w^*)xx^* = x^*=\gamma(v)^*. 
\end{equation}
Because $\psi_i=d(\theta)^{\frac12}\cdot w^{i*}v$ by definition of
$v$, $v$ and $N$ generate $M$, so $\gamma$ is indeed an endomorphism. 
Hence $(\gamma,v,w)$ is a Q-system, and $(\theta,w,x)$ its dual
($\theta=\gamma\rest_N$ and $x=\gamma(v)$). Finally $N=w^*\gamma(M)w$
because $w^*\gamma(vnv^*)w=d(\theta)\inv\cdot n$, showing that the
Q-systems are indeed the Q-system associated with $N\subset M$ and its
dual. The subfactor is irreducible since $\id$ is contained in
$\theta$ with multiplicity one.

(iii) is now obvious. \QED

\section{$\alpha$-induction}
\setcounter{equation}{0}
We collect a number of well-known results on $\alpha$-induction
\cite{LR,BE1}, used in the course of our arguments in this article.

{\bf B.1 Definition \cite{LR}:} Let $N$ be a factor, and $\Delta$ a set
of endomorphisms $\varrho$ of $N$ equipped with a braiding
$\eps(\varrho_1,\varrho_2):\varrho_1\varrho_2\to\varrho_2\varrho_1$, giving 
rise to a braided C* tensor category with direct sums and subobjects. 
Let $N\subset M$ be an irreducible subfactor with canonical
endomorphism $\gamma$, and dual canonical endomorphism
$\theta=\gamma\rest_N$ such that $\theta\in\Delta$. Then for
$\varrho\in\Delta$,  
\begin{equation}
\alpha^+_\varrho:=\gamma\inv\circ \Ad_{\eps(\theta,\varrho)}
\circ\varrho\circ\gamma 
\end{equation}
extends the endomorphism $\varrho\in\Delta$ of $N$ to an endomorphism
$\alpha^+_\varrho$ of $M$. (The nontrivial fact is that  
$\Ad_{\eps(\theta,\varrho)} \circ\varrho\circ\gamma (M)$ belongs to
$\gamma(M)$.) One has  
\begin{equation}
\alpha^+_\varrho(n)= \varrho(n)\qquad\hbox{and}\qquad 
\alpha^\pm_\varrho(v)=\eps(\theta,\varrho)v
\end{equation}
for $n\in N$ and $v$ the canonical isometry in the Q-system $(\gamma,v,w)$. 
The same holds true for $\alpha^-_\varrho$, replacing the braiding in 
(B.1), (B.2) with the opposite braiding
$\eps^-(\varrho_1,\varrho_2)=\eps(\varrho_2,\varrho_1)^*$. The endomorphisms
$\alpha^\pm_\varrho$ are invariant under inner conjugations of the
Q-system (i.e., $\gamma\mapsto \Ad_u\circ \gamma$, $v\mapsto uv$, $w\mapsto
uw$, $u\in N$ unitary). 

{\bf B.2 Proposition \cite{LR}:} If $I\mapsto[A(I)\subset B(I)]$ is a 
(chiral) quantum field theoretical net of subfactors, then the dual
canonical endomorphism of $A(I)$ for $A(I)\subset B(I)$
extends to a DHR endomorphism $\theta$ of the net $A$ with $[\theta]$
independent of $I$. With $\Delta$ the set of DHR endomorphisms and
$\eps(\rho_1,\rho_2)$ the DHR braiding \cite{FRS1}, $\alpha^\pm_\rho$
can be defined as endomorphisms of the net $B$ such that (B.2) and its
analog for $\alpha^-_\rho$ still hold. If $\rho$ is localized in $I$,
then $\alpha^+_\rho$ (resp.\ $\alpha^-_\rho$) is a semi-localized
endomorphism of $B$, i.e., it acts trivially on $B(K)$ whenever $I<K$
(resp.\ $I>K$).    

{\bf B.3 Proposition:} Let $\rho$, $\sigma$ be localized in $I$. Then
for every combination of $\eps=\pm$, $\eps'=\pm$, the space of 
global intertwiners 
\begin{equation}
\{t\in B: t\alpha^\eps_\rho(b)=\alpha^{\eps'}_\sigma(b)t \; \hbox{for
  all}\; b\in B\}
\end{equation}
coincides with the space of local intertwiners
\begin{equation}
\{t\in B(I): t\alpha^\eps_\rho(b)=\alpha^{\eps'}_\sigma(b)t \; 
\hbox{for all}\; b\in B(I)\}.
\end{equation}

{\it Proof:} Every element $b\in B$ can be written uniquely as $b=av$
with $a\in A$ and $v\in B(I)$ the charged intertwiner of the Q-system
$(\gamma,v,w)$ for $A(I)\subset B(I)$ \cite{LR}; furthermore $b\in
B(I)$ iff $a\in A(I)$. Let $t=av$ be a global intertwiner. Then 
$t\alpha^\eps_\rho(a_1)=\alpha^{\eps'}_\sigma(a_1)t$ implies that $a$ is
a global intertwiner in $\Hom(\theta\rho,\sigma)$, hence $a\in A(I)$
by Haag duality of $A$, hence $t\in B(I)$ is in fact a local intertwiner. 

Conversely, let $t=av$ be a local intertwiner. Then $a$ is local
intertwiner, hence \cite[Thm.\ 2.3]{GL} a global intertwiner in
$\Hom(\theta\rho,\sigma)$. Thus, we have
$t\alpha^\eps_\rho(b)=\alpha^{\eps'}_\sigma(b)t$ for 
all $b\in B(I)$ by assumption, and for all $b\in A(I')$ by the trivial
action of the endomorphisms on $A(I')$ and locality. Since $B(I)$ and
$A(I')$ generate all of $B$ by strong additivity of $A$, $t$ is a
global intertwiner. \QED

{\bf B.4 Proposition \cite[Lemma 3.5 and Lemma 3.25]{BE1}:} Let
$\rho$, $\sigma$, $\tau$ be DHR endomorphisms of $A$. 

(i) If $T\in\Hom(\rho,\sigma)\subset A$, then also   
$T\in\Hom(\alpha^\pm_\rho,\alpha^\pm_\sigma)\subset B$. 

(ii) If $t \in\Hom(\alpha^\pm_\rho,\alpha^\pm_\sigma)$, then the
naturality relations $\alpha^\pm_\tau(t)\eps(\rho,\tau)=
t\eps(\sigma,\tau)$ and $\alpha^\pm_\tau(t)\eps(\tau,\rho)^*=
t\eps(\tau,\sigma)^*$ hold.  
 
\vskip5mm
{\bf Acknowledgements.} The second author (KHR) thanks the
Dipartimento di Matematica of the University of Rome ``Tor Vergata''
(where the idea for this work was created) and the Max-Planck
Institute for Physics in Munich (where part of the work was completed)
for hospitality, D. Evans and Y. Kawahigashi for helpful
correspondence, and J. Fuchs for interesting discussions.

\end{document}